\documentclass[pra,twocolumn,showpacs,preprintnumbers,amsmath,amssymb]{revtex4-1}
\usepackage{amssymb,amsmath}
\usepackage{color}
\usepackage{graphicx}
\usepackage{dcolumn}
\usepackage{bm}
\usepackage{latexsym,epsfig}

\begin{document}
\preprint{\today}

\title{Random-Phase-Approximation Excitation Spectra for Bose-Hubbard Models}

\author{Jamshid Moradi Kurdestany}
\email{jamshid@physics.iisc.ernet.in} \affiliation{Jawaharlal Nehru Centre For Advanced Scientific Research,
Jakkur, Bangalore, India and Department of Physics and Astronomy,
University of Missouri,
Columbia, MO 65211-7010, USA}\altaffiliation[Also
at~]{Centre for
Condensed Matter Theory, Department of Physics, Indian Institute of
Science, Bangalore 560 012, India}
\author{Ramesh V. Pai}
\email{rvpai@unigoa.ac.in} \affiliation{Department of Physics, Goa
University, Taleigao Plateau, Goa 403 206, India}
\author{Rahul Pandit}
\email{rahul@physics.iisc.ernet.in} \altaffiliation[Also
at~]{Jawaharlal Nehru Centre For Advanced Scientific Research,
Jakkur, Bangalore, India} \affiliation{Centre for Condensed Matter
Theory, Department of Physics, Indian Institute of Science,
Bangalore 560012, India.}
\date{\today}
\begin{abstract}
We obtain the excitation spectra of the following three generalized
Bose-Hubbard (BH) models: (1) a two-species generalization of the spinless BH
model, (2) a single-species, spin-1 BH model, and (3) the extended Bose-Hubbard
model (EBH) for spinless interacting bosons of one species. In all the phases
of these models we provide a unified treatment of random-phase-approximation
(RPA) excitation spectra.  These spectra have gaps in all the MI phases and
gaps in the DW phases in the EBH model; they are gapless in all the SF phases
in these models and in the SS phases in the EBH model. We obtain the dependence
of (a) gaps $\Delta$ and (b) the sound velocity $u_s$ on the parameters of
these models and examine $\Delta$ and $u_s$ as these systems go through phase
transitions.  At the SF-MI transitions in the spin-1 BH model, $u_s$ goes to
zero continuously (discontinuously) for MI phases with an odd (even) number of
bosons per site; this is consistent with the natures of these transition in
mean-field theory.  In the SF phases of these models, our excitation spectra
agree qualitatively, at weak couplings, with those that can be obtained from
Gross-Pitaevskii-type models. We compare the results of our work with earlier
studies of related models and discuss implications for experiments.
\end{abstract}

\maketitle

\section{Introduction}

Cold-atom systems, such as spin-polarized $^{87}$Rb in traps,
have provided us excellent laboratories for studies of quantum
phase transitions~\cite{rmp55,sachdev55}. One example of such a
transition is the superfluid (SF) to bosonic Mott-insulator (MI)
transition~\cite{jaksch55,greiner55}.  This SF-MI transition,
first predicted by mean-field studies~\cite{fisher55,sheshadri55}
and later investigated in Monte-Carlo simulations~\cite{mc55} of the
Bose-Hubbard (BH) model, has been obtained in experiments in
systems of interacting bosons in optical
lattices~\cite{rmp55,jaksch55,greiner55}.

Experimental realizations have also been found for generalized BH
models. For instance, recent
experiments~\cite{catani55,trotzky0855,Widera55,Weld55,Gadway55}
on a degenerate mixture of two types of bosons, namely, $^{87}$Rb
and $^{41}$K, in a three-dimensional optical lattice, have
yielded a laboratory realization of a BH model with two species
of interacting bosons, which has been studied
theoretically~\cite{kuklov0355,han0455,buonsante0855,hu0955,ozaki0955,JMK12,rvpai1255}
and by Monte Carlo simulations~\cite{roscilde0755}. Systems of
cold alkali atoms, with nuclear spin $I=3/2$ and a hyperfine spin
$F=1$, such as, $^{23}$Na, $^{39}$K, and $^{87}$Rb, in purely
optical (and not magnetic) traps, can lead to realizations of the
spin-1 Bose-Hubbard model, with spinor
condensates~\cite{spin1expt55,tlho55,mukerjee,rvpspin155}, which
can exhibit polar and ferromagnetic superfluid phases in addition
to MI phases. A dipolar condensate of $^{52}{\rm Cr}$
atoms~\cite{werner05} has been obtained; to model this we must
include long-range interactions~\cite{goral02} in the BH model,
in addition to the repulsive interaction between bosons on the
same lattice site.  The first step in this direction can be taken
by studying the extended-Bose-Hubbard (EBH) model, which has SF,
MI, density-wave (DW), and supersolid (SS)
phases~\cite{kovrizhin0544,andreev69,kim04,JMK12}.

The study of Ref.\cite{sheshadri55} has introduced a simple,
transparent, mean-field theory, which yields the phases of the BH
model; this study has also developed a random-phase-approximation
(RPA) calculation, which builds upon their mean-field theory, to
obtain the excitation spectra in the phases of the BH model. We
generalize such RPA calculations so that they can be used for the
other bosonic models, namely, the extended-Bose-Hubbard (EBH)
model, the spin-1 Bose-Hubbard model, and the Bose-Hubbard model
with two types of bosons. We then use this RPA to obtain the
excitation spectra in all the phases of these generalized BH
models, the mean-field theories for which we have discussed in Refs.\cite{JMK12,rvpai1255}.

Our main goal is to provide a unified treatment of RPA excitation
spectra in the generalized BH models we have mentioned above.
There have been a few studies of excitation spectra in some of
the phases of these models; not all of them are formulated in the
way we describe here. We discuss the relation of our work with
other studies in the last Section of this paper.

In addition to providing a unified treatment of RPA excitation
spectra for the three BH models above, our study yields several
interesting results, which we summarize below, before we proceed to
the details of our work. Our RPA yields excitation spectra, which
have gaps in all the MI phases, in all these BH models, and gaps
in the DW phases in the EBH model. These spectra are gapless in
all the SF phases in these models and in the SS phases in the EBH
model. We obtain the dependence of (a) gaps $\Delta$ and (b) the
sound velocity $u_s$ on the parameters of these models. In
particular, we examine $\Delta$ and $u_s$ as these systems go
through phase transitions.  We find, e.g., that, at the SF-MI
transitions in the spin-1 BH model, $u_s$ goes to zero
continuously (discontinuously) for MI phases with an odd (even)
number of bosons per site; this is consistent with the natures of
these transition (continuous or discontinuous) in the mean-field
theory for the spin-1 BH model~\cite{rvpspin155}. In the SF phases of these
models, our excitation spectra agree qualitatively, at weak
couplings, with those that can be obtained from
Gross-Pitaevskii-type models.  For example, our excitation
spectra are qualitatively similar to those obtained for
ferromagnetic and polar superfluids in ~\cite{tlho55}, which
uses a spin-1 generalization of the Gross-Pitaevskii equation.

The remaining part of this paper is organized as follows. In
Sec.~\ref{sec:models} we define the models we study, give the
elements of the mean-field theory that we use to obtain the
phases of these models, and then show how to develop, and then
close at the level of the RPA, the equations of motion for the
Green functions for all these models.  In Sec.~\ref{sec:results}
we present our plots of RPA excitation spectra for representative
parameter values in these BH models. The concluding Section
contains a discussion of our results and a comparison of these
with earlier studies of such excitation spectra.

\section{Models, Mean-field Theory, and the Random-Phase Approximation
}\label{sec:models}

In the first subsection~\ref{subsec:models} below we define the
BH model with two species of bosons, the spin-1 BH model, and the
EBH model. The next subsection~\ref{subsec:Mean-field theory}
outlines our mean-field theories for these models, principally to
establish notations that are required for the development of the
RPA, which we present in the last subsection~\ref{subsec:RPA
Excitation Spectra}. Details of our mean-field theories have been
discussed in~\cite{sheshadri55,JMK12,rvpai1255}.

\subsection{Models}\label{subsec:models}
A Bose-Hubbard (BH) model, with two types of bosons, is defined by
the following Hamiltonian:
\begin{eqnarray}
\label{eq:bhab}\frac{{\cal H}}{z} &=& -\frac{t_a}{z} \sum_{<i,j>}
(a_{i}^{\dagger} a_{j} + h.c.)-\frac{t_b}{z} \sum_{<i,j>}
(b_{i}^{\dagger} b_{j} + h.c.)\nonumber \\
&&+\frac{1}{2}\frac{U_a}{z} \sum_{i} {\hat n}_{ai} ({\hat n}_{ai}
-1)+\frac{1}{2}\frac{U_b}{z} \sum_{i} {\hat n}_{bi} ({\hat n}_{bi}
-1)\nonumber \\
&&+\frac{U_{ab}}{z} \sum_{i} {\hat n}_{ai}{\hat
n}_{bi}-\frac{\mu_{a}}{z}\sum_{i} {\hat n}_{ai}
-\frac{\mu_{b}}{z}\sum_{i} {\hat n}_{bi};
\end{eqnarray}
the first and second terms represent, respectively, the hopping
of bosons of types $a$ and $b$ between the nearest-neighbor pairs
of sites $<i,j>$, with hopping amplitudes $t_a$ and $t_b$; here
$a_{i}^\dagger,\, a_{i},$ and ${\hat n}_{ai}\equiv
a^{\dagger}_{i}a_{i}$ and $b_{i}^\dagger,\, b_{i},$ and ${\hat
n}_{bi}\equiv b^{\dagger}_{i}b_{i}$ are, respectively, boson
creation, annihilation, and number operators at the sites $i$ of
a $d$-dimensional hypercubic lattice (we present excitation
spectra for $d=2$) for the two bosonic species.  For simplicity,
we restrict ourselves to the case $t_a = t_b = t$, and, to set
the scale of energy, we use $zt = 1 $, where $z=2d$ is the
nearest-neighbor coordination number. The third and fourth terms
account for the onsite interactions of bosons of a given type,
with energies $U_a$ and $U_b$, respectively, whereas the fifth
term, with energy $U_{ab}$, arises because of the onsite
interactions between bosons of types $a$ and $b$.  We have two
chemical-potential terms, $\mu_a$ and $\mu_b$, which control,
respectively, the total number of bosons of species $a$ and $b$.

We also study the following spin-1 BH Hamiltonian~\cite{rvpspin155} on
a $d-$dimensional hypercubic lattice with sites $i$:
\begin{eqnarray}
\label{eq:bhspin1} {\frac{\cal{H}}{zt}}=&& -
\sum_{<i,j>,\sigma}(a^{\dagger}_{i,\sigma}a_{j,\sigma}+ h.c.) +
\frac{1}{2}\frac{U_0}{zt} \sum_i {\hat n}_i({\hat{n}_i}-1)
\nonumber \\
&&+\frac{1}{2}\frac{U_2}{zt} \sum_i (\vec{F}^2_i -
2\hat{n_i})-\frac{\mu}{zt}\sum_{i} \hat{n}_i,
\end{eqnarray}
where spin-1 bosons hop between the nearest-neighbor pairs of
sites $<i,j>$ with amplitude $t$, the spin index $\sigma$ can
be $1,0,-1$, $a^{\dagger}_{i,\sigma}$ and $a_{i,\sigma}$ are,
respectively, site- and spin-dependent boson creation and
annihilation operators, and the number operator
$\hat{n}_{i\sigma}\equiv a^{\dagger}_{i,\sigma}a_{i,\sigma}$; the
total number operator at site $i$ is
$\hat{n}_i\equiv\sum_\sigma\hat{n}_{i,\sigma}$, and
$\vec{F}_i=\sum_{\sigma,\sigma '} a^{\dagger}_{i,\sigma}
\vec{F}_{\sigma ,\sigma '} a_{i,\sigma '}$, with $ \vec{F}_{\sigma
,\sigma '}$ standard spin-1 matrices. This
model~(\ref{eq:bhspin1}) includes, in addition to the onsite
repulsion $U_0$, an energy $U_2$, for nonzero spin configurations
on a site, which arises from the difference between the
scattering lengths for $S=0$ and $S=2$ channels~\cite{law55}. The
chemical potential $\mu$ controls the total number of bosons.

The Hamiltonian for the EBH model is
\begin{eqnarray}
\frac{{\cal H}}{zt} &=& -\frac{1}{z} \sum_{<i,j>} (a_{i}^{\dagger}
a_{j} + h.c.)+\frac{1}{2}\frac{U}{zt} \sum_{i} {\hat n}_{i} ({\hat
n}_{i} -1)
\nonumber \\
&+&\frac{V}{zt} \sum_{<i,j>}{\hat n}_i{\hat n}_{j}- \frac{\mu}{zt}
\sum_i{\hat n}_i , \label{eq:ebh}
\end{eqnarray}
where $t$ is the amplitude for a boson to hop from site $i$ to its
nearest-neighbor site $j$, $z$ is the nearest-neighbor coordination
number, $<i,j>$ are nearest-neighbor pairs of sites, $h.c.$ denotes
the Hermitian conjugate, $a_i^{\dag}, \, a_i$, and ${\hat n}_i
\equiv a_i^{\dag} a_i$ are, respectively, boson creation,
annihilation, and number operators at the site $i$, the repulsive
potential between bosons on the same site is $U$, the chemical
potential $\mu$ controls the total number of bosons, and $V$
is the repulsive interaction between bosons on nearest-neighbor sites.\\


To make a detailed comparison of our results with experiments, the
parameters of the BH models must be related to experimental
ones~\cite{rmp55}. For the simple BH model this is done as follows:
$\frac{U}{zt}=\frac{\sqrt{8}\pi}{4z}
\frac{a_s}{a}\exp({2\sqrt{\frac{V_0}{E_r}}})$, where $E_r$ is the
recoil energy, $V_0$ the strength of the lattice potential, $a_s$
($=5.45$~nm for $^{87}$Rb) the $s$-wave scattering coefficient,
$a=\lambda/2$ the optical-lattice constant, and $\lambda=825$~nm the
wavelength of the laser used to create the optical-lattice;
typically $0 \leq V_0 \leq 22 E_r$. If we use this experimental
parametrization, we scale all the energies by $E_r$. (In this paper,
we set $zt=1$, i.e., we measure all energies in units of $zt $.)

A two-species BH model, has been realized in an optical lattice by
using elliptically polarized light. By changing the polarization
angle it is possible to shift the lattices with respect to each
other, and, thereby, control the interactions between the two
species of bosons and also their hopping amplitudes (see
Refs.~\cite{Mandel2003, Jaksch1999, Brennen1999} for details).

The spin-dependent term in the spin-1 BH model follows from the
difference between the scattering lengths $a_0$ and $a_2$, for
$S=0$ and $S=2$ channels~\cite{law55}, respectively.  These lengths
yield $U_0=4\pi\hbar^2(a_0+2 a_2)/3M$ and $U_2=4\pi\hbar^2(a_2-
a_0)/3M$, where $M$ is the mass of the atom~\cite{tlho55}. If we
consider $^{23}$Na, $a_2=54.7 a_B$ and $a_0=49.4 a_B$, where
$a_B$ is the Bohr radius, so $U_2 > 0$; in contrast, for
$^{87}$Rb, $a_2=(107\pm 4) a_B$ and $a_0=(110\pm 4) a_B$, so $U_2
$ can be negative.

For the EBH case, the relation of our parameters to those in
dipolar bose systems~\cite{werner05,goral02} is not straightforward
because of long-range interactions, but we can use the
following estimates:
\begin{equation}\label{eq:t}
    t=\int
    w^*({\bf r}-{\bf r}_i)[\frac{-\hbar^2}{2m}{\nabla^2}+V_l({\bf r})]w({\bf r}-{\bf r}_j)d^3r ,
\end{equation}
where $i$ and $j$ are nearest-neighbor sites, $w$ are Wannier
functions, and $ V_l({\bf r})=\sum_{\alpha=x,y,z}V^2_\alpha
\cos^2(k_\alpha \alpha)$ is the optical-lattice potential with
wavevector $\bf{k}$. Furthermore,
\begin{equation}\label{eq:U}
U=U_{ii}=\int |w({\bf r}- {\bf r}_i)|^2 V_{\rm int}({\bf r}-{\bf
r'}) |w({\bf r'}- {\bf r}_i)|^2 d^3r~ d^3{r'}
\end{equation}
and
\begin{equation}\label{eq:V}
V=U_{<ij>}=\int |w({\bf r}- {\bf r}_i)|^2 V_{\rm int}({\bf r}-{\bf
r'}) |w({\bf r'}- {\bf r}_j)|^2 d^3r~ d^3{r'},
\end{equation}
with
\begin{equation}\label{eq:VINT}
V_{\rm int}=D^2 \frac{1-3\cos^2\theta}{|{\bf r}-{\bf r'}|^3} +
\frac{4\pi\hbar^2a_s}{M}\delta({\bf r}-{\bf r'}),
\end{equation}
where $D$ is the dipole moment, $a_s$ is the $s$-wave  scattering
constant, and $M$ is the mass of the atom. The $s$-wave
scattering constant of Chromium is $|a(^{52}{\rm
Cr})|=(170\pm39)a_0$ and $|a(^{50}{\rm Cr})|=(40\pm15)a_0$, where
$a_0=0.053$~nm~\cite{chrom03}.

\subsection{Mean-field theory}\label{subsec:Mean-field theory}

We use a homogeneous mean-field theory for these models because
we do not include a quadratic confining potential.  (With such a
potential, we must use an inhomogeneous version of this
mean-field theory~\cite{JMK12,rvpai1255}.) Conventional
mean-field theories introduce a decoupling approximation that
reduces a model with interacting bosons or fermions to an
effective, noninteracting problem, which can be solved easily
because the effective Hamiltonian is quadratic in boson or
fermion operators. By contrast, the mean-field theories of
~\cite{sheshadri55,JMK12,rvpai1255} decouple the hopping terms, in
the BH models defined above; these hopping terms are quadratic in
boson operators, so, after this decoupling, we obtain effective,
one-site Hamiltonians, which can be diagonalized numerically.

For the two-species BH model Eq.(\ref{eq:bhab}), our mean-field theory
~\cite{rvpai1255} obtains an effective,
one-site problem by decoupling the two hopping terms as follows:
\begin{eqnarray}
a^{\dagger}_{i}a_{j} &\simeq& \langle a^{\dagger}_{i}\rangle
a_{j} +a^{\dagger}_{i}\langle a_{j}\rangle -\langle a^{\dagger}_{i}\rangle
\langle a_{j}\rangle ; \nonumber \\
b^{\dagger}_{i}b_{j} &\simeq& \langle b^{\dagger}_{i}\rangle
b_{j} +b^{\dagger}_{i}\langle b_{j}\rangle -\langle b^{\dagger}_{i}\rangle
\langle b_{j}\rangle ;
\label{eq:hopab}
\end{eqnarray}
the superfluid order parameters for the site $i$ for
bosons of types $a$ and $b$ are
$\psi_{{ai}}\equiv \langle a_{i}\rangle$ and $\psi_{{bi}}\equiv \langle b_{i}\rangle$, respectively.
The approximation Eq.(\ref{eq:hopab}) can now be used to write the
Hamiltonian~(\ref{eq:bhab}) as a sum over single-site, mean-field
Hamiltonians ${\cal{H}}_i^{MF}$ that are given below:
\begin{eqnarray}
 \frac{{\cal{H}}_i^{MF}}{zt}&=&\frac{1}{2}\frac{U_a}{zt}
 {\hat{n}}_{ai}({\hat{n}}_{ai}-1)
-\frac{\mu_{a}}{zt} {\hat{n}}_{ai}\nonumber
\\&-&(\phi_{ai}a^{\dagger}_{i}
+\phi_{ai}^*a_{i})+
\psi_{ai}^* \phi_{ai}\nonumber \\
&+&\frac{1}{2}\frac{U_b}{zt} {\hat{n}}_{bi}({\hat{n}}_{bi}-1)
-\frac{\mu_{b}}{zt} {\hat{n}}_{bi}\nonumber
\\&
-&(\phi_{bi}b^{\dagger}_{i}+\phi_{bi}^*b_{i})+ \psi_{bi}^*
\phi_{bi}\nonumber
\\&+&\frac{U_{ab}}{z} {\hat{n}}_{ai}{\hat{n}}_{bi}.
\label{eq:mfhab}
\end{eqnarray}
Here, $\phi_{ai}\equiv \frac{1}{z}\sum_{\delta} \psi_{{ai}+\delta}$ and
$\phi_{bi}\equiv \frac{1}{z}\sum_{\delta} \psi_{{bi}+\delta}$, where $\delta$
labels the nearest neighbors of the site $i$.  For the homogeneous case,
the onsite chemical potentials are $\mu_{a}$ and $\mu_{b}$, for all
$i$, so $\rho_{ai}=\rho_{a}$, $\rho_{bi}=\rho_{b}$, $\psi_{ai}=\psi_{a}$, and
$\psi_{bi}=\psi_{b}$ are independent of $i$.

The mean-field theory for the spin-1 BH model follows
along similar lines~\cite{rvpspin155}:
\begin{equation}
a^{\dagger}_{i,\sigma}a_{j,\sigma} \simeq \langle
a^{\dagger}_{i,\sigma}\rangle a_{j,\sigma} +
a^{\dagger}_{i,\sigma}\langle a_{j,\sigma}\rangle - \langle
a^{\dagger}_{i,\sigma}\rangle \langle a_{j,\sigma}\rangle
\label{eq:hopspin1}
\end{equation}
and
\begin{eqnarray}
 \frac{{\cal{H}}_i^{MF}}{zt}&=&\frac{1}{2}\frac{U}{zt} {\hat{n}_i}({\hat{n}_i}-1)
+ \frac{1}{2}\frac{U_2}{zt} ({\vec F}^2_i-2{\hat{n}_i})
-\frac{\mu}{zt} {\hat{n}_i} \nonumber
\\ &-&\sum_\sigma(\phi_{i,\sigma}a^{\dagger}_{i,\sigma}+
\phi_{i,\sigma}^*a_{i,\sigma})+ \sum_\sigma\psi_{i,\sigma}^*
\phi_{i,\sigma}. \label{eq:mfhspin1}
\end{eqnarray}
Here, we use the following superfluid order parameters:
\begin{equation}
\psi_{i,\sigma}\equiv \langle a^{\dagger}_{i,\sigma}\rangle \equiv
\langle a_{i,\sigma}\rangle;
\label{eq:opspin1}
\end{equation}
and $\phi_{i,\sigma}\equiv \frac{1}{z}\sum_{\delta} \psi_{(i+\delta),\sigma}$,
where $\delta$ labels the $z$ nearest neighbors of the site $i$; recall,
furthermore, that $\sigma$ can assume the values $1,\,0,\,-1$, and
$\hat{n}_{{i\sigma}}\equiv a^{\dagger}_{i,\sigma}a_{i,\sigma}$,
$\hat{n}_{i}\equiv\sum_\sigma\hat{n}_{{i,\sigma}}$, and
$\vec{F}_i=\sum_{\sigma,\sigma '} a^{\dagger}_{i,\sigma} \vec{F}_{\sigma
,\sigma '} a_{i,\sigma '}$, with $ \vec{F}_{\sigma ,\sigma '}$ standard spin-1
matrices. For the homogeneous case, the chemical potential $\mu$ does
not depend on $i$, so $\rho_{i,\sigma}=\rho_\sigma$ and $\psi_{i,\sigma}=\psi_\sigma$ are also
independent of $i$.

In the EBH model with $V > 0$, we decouple the first and third
terms of Eq.(\ref{eq:ebh}) to obtain an effective one-site
problem~\cite{JMK12}, which neglects quadratic deviations from equilibrium
values (denoted by angular brackets).  The two approximations we
use are as follows:
\begin{eqnarray} a^{\dagger}_{i}a_{j} &\simeq& \langle a^{\dagger}_{i}\rangle
a_{j} +a^{\dagger}_{i}\langle a_{j}\rangle -\langle a^{\dagger}_{i}\rangle
\langle a_{j}\rangle ; \nonumber \\
{\hat n}_{i}{\hat n}_{j} &\simeq& \langle
{\hat n}_{i}\rangle {\hat n}_{j} +{\hat n}_{i}\langle {\hat n}_{j}\rangle
-\langle {\hat n}_{i}\rangle \langle {\hat n}_{j}\rangle;
\label{eq:decoup}
\end{eqnarray}
the superfluid order parameter and the local density for
the site $i$ are, respectively, $\psi_{i}\equiv \langle a_{i}\rangle$ and
$\rho_{i} \equiv \langle {\hat n}_{i}\rangle $.  The
approximation~(\ref{eq:decoup}) can now be used to write the
Hamiltonian~(\ref{eq:ebh}) as a sum over single-site, mean-field
Hamiltonians ${\cal{H}}_i^{MF}$ as follows:
\begin{eqnarray}
{\cal{H}}^{MF} &\equiv& \sum_{i} {\cal{H}}_i^{MF} , \nonumber \\
\frac{{\cal{H}}_i^{MF}}{zt} &\equiv&
\frac{1}{2}\frac{U}{zt} {\hat{n}}_i({\hat{n}}_i-1) -\frac{\mu}{zt}
{\hat{n}}_i -(\phi_ia^{\dagger}_{i}+\phi_i^*a_{i}) \nonumber \\
&+&\frac{1}{2}(\psi_i^* \phi_i+\psi_i \phi_i^*)+
{\frac{V}{t}}({\hat{n}}_i{\bar{\rho}}_i-{\rho_i}{\bar{\rho}}_i),
\label{eq:mfham}
\end{eqnarray}
where the superscript $MF$ stands for mean-field, and $\phi_i\equiv
\frac{1}{z}\sum_{\delta} \psi_{i+\delta}$ , ${\bar{\rho}}_i\equiv
\frac{1}{z}\sum_{\delta} \rho_{i+\delta}$, and $\delta$ labels the
$z$ nearest neighbors of the site $i$. We can have density-wave (DW)
and supersolid (SS) phases, so our order parameters should allow for
such phases. In the hypercubic lattices we consider, there are two
sublattices ${\mathcal A}$ and ${\mathcal B}$. Each site on the
${\mathcal A}$ (${\mathcal B}$) sublattice has $z$ nearest
neighbors, each one of which belongs to the ${\mathcal B}$
(${\mathcal A}$) sublattice; thus, $\psi_{i}=\psi_A$ and $\bar{
\rho}_{i}=\rho_A$, if $i\in {\mathcal A}$, and $\psi_{i}=\psi_B$ and
$\bar{\rho}_{i}=\rho_B$, if $i\in {\mathcal B}$, whereas
$\phi_{i}=\psi_B$ and $\bar{\rho}_{i}=\rho_B$, if $i\in {\mathcal
A}$, and $\phi_{i}=\psi_A$ and ${\bar \rho}_{i}=\rho_A$, if $i\in
{\mathcal B}$.  The chemical potentials, which are conjugate to
$\rho_A$ and $\rho_B$, respectively, are $\mu_A$, if $i\in {\mathcal
A}$, and $\mu_B$, if $i\in {\mathcal B}$; similarly, we can define
creation, annihilation, and number operators for each sublattice.
The mean-field Hamiltonian~(\ref{eq:mfham}) can now be written as
\begin{equation}
\cal{H}_{AB}^{MF} \equiv \cal{H}_{A}^{MF}+\cal{H}_{B}^{MF},
\label{eq:hab1}
\end{equation}
where
\begin{eqnarray}
\frac{{\cal{H}}_A^{MF}}{zt} &\equiv&
-(a_{A}\psi_B^*+a^{\dagger}_{A}\psi_B)+
\frac{1}{2}(\psi_A\psi_B^*+\psi_A^*\psi_B)
\nonumber \\
&+& \frac{V}{t}({\hat{n}}_A{\rho}_B-{\rho}_A{\rho}_B)+
\frac{1}{2}\frac{U}{zt}{{\hat{n}}_A}({{\hat{n}}_A}-1)\nonumber\\
&-&\frac{\mu_A}{zt} {\hat{n}}_A; \label{eq:hab2}
\end{eqnarray}

\begin{eqnarray}
\frac{{\cal{H}}_B^{MF}}{zt} &\equiv&
-(a_{B}\psi_A^*+a^{\dagger}_{B}\psi_A)+
\frac{1}{2}(\psi_B\psi_A^*+\psi_B^*\psi_A)
\nonumber \\
&+& \frac{V}{t}({\hat{n}}_B{\rho}_A-{\rho}_B{\rho}_A)+
\frac{1}{2}\frac{U}{zt}{\hat{n}}_B({\hat{n}}_B-1)\nonumber
\\
&-&\frac{\mu_B}{zt} {\hat{n}}_B. \label{eq:hab3}
\end{eqnarray}

\subsection{Random-Phase Approximation (RPA) Excitation Spectra}\label{subsec:RPA Excitation Spectra}

We now present a systematic method for developing the RPA for generalized BH
models of the types discussed above. Our RPA is based on that of Ref.~\cite{sheshadri55}, for the simple BH model, which is in turn a
generalization of the work of Ref.~\cite{Haley} for the spin model. Such RPA calculations use the mean-field order parameters defined
above and the eigenstates of our
mean-field Hamiltonians of Eqs.(\ref{eq:mfhab}), (\ref{eq:mfhspin1}), and (\ref{eq:hab1}).

We begin by defining the projection operators~\cite{Haley}
$L_{\alpha\alpha'}^{i}=|i\alpha\rangle \langle i\alpha'|$, where $i$ is the
site index and $|i\alpha\rangle$ are the eigenstates of the mean-field
Hamiltonian. Any single-site operator $\hat{O}$ can be expressed as $\hat{O}
\equiv\sum_{\alpha\alpha'}\langle i\alpha|\hat{O}|i\alpha'\rangle
L_{\alpha\alpha'}^{i}$. In the next three subsections, we obtain, explicitly,
the equations of motion for the Green functions for the three generalized BH
models we consider and show how to close them in the RPA (see the
Appendix for details).

\subsubsection{RPA Excitation Spectra for the two-species BH model}

If we use $\hat{O} \equiv\sum_{\alpha\alpha'}\langle
i\alpha|\hat{O}|i\alpha'\rangle L_{\alpha\alpha'}^{i}$, Eq.(\ref{eq:bhab})
becomes (see the Appendix)
\begin{equation}
\label{eq:RPAES2BHh}
{\cal{H}} = -\sum_{i\alpha\alpha'} V^{i}_{\alpha\alpha'}
L^{i}_{\alpha\alpha'}-\sum_{<i,j>,\alpha\alpha',\beta\beta'}
T^{ij}_{\alpha\alpha',\beta\beta'}
L^{i}_{\alpha\alpha'}L^{j}_{\beta\beta'},
\end{equation}
where
\begin{eqnarray}
\label{eq:RPAES2BHv}
\nonumber
V^{i}_{\alpha\alpha'}&=&(\mu_{a}+\frac{U_a}{2})\langle
i\alpha|\hat{n}_a|i\alpha'\rangle -\frac{U_a}{2}\langle
i\alpha|\hat{n}_a^{2}|i\alpha'\rangle \\
\nonumber
&&+(\mu_{b}+\frac{U_b}{2})\langle i\alpha|\hat{n}_b|i\alpha'\rangle
-\frac{U_b}{2}\langle i\alpha|\hat{n}_b^{2}|i\alpha'\rangle\\
 &&-U_{ab}\langle i\alpha|\hat{n}_a\hat{n}_b|i\alpha'\rangle
\end{eqnarray}
and
\begin{eqnarray}
\label{eq:RPAES2BHt}
\nonumber
T^{ij}_{\alpha\alpha',\beta\beta'} &=& \langle
i\alpha|\hat{a}^{\dagger}|i\alpha'\rangle \langle
j\beta|\hat{a}|j\beta'\rangle+\langle
i\alpha|\hat{a}|i\alpha'\rangle
\langle j\beta|\hat{a}^{\dagger}|j\beta'\rangle\\
\nonumber &&+\langle i\alpha|\hat{b}^{\dagger}|i\alpha'\rangle \langle
j\beta|\hat{b}|j\beta'\rangle+\langle
i\alpha|\hat{b}|i\alpha'\rangle \langle
j\beta|\hat{b}^{\dagger}|j\beta'\rangle.\\
&&~
\end{eqnarray}

In the RPA, averages of products of operators
are replaced by products of their averages (at finite temperature we have
thermal averages, but we restrict ourselves to zero temperature here),
so the equations of motion for the Green functions, for this BH model with
two species of bosons, namely,
\begin{eqnarray}
\label{eq:RPAES2BHg}
G^{ij}_{\alpha\alpha', \beta\beta'}(t,t') = -i\theta(t-t')\langle[L^{i}_{\alpha\alpha'}(t);
 L^{j}_{\beta\beta'}(t')]\rangle
\end{eqnarray}
become, after
Fourier transforms over space and time, the following
linear equations, which can be inverted easily:
 \begin{eqnarray}
\label{eq:RPAES2BHw}
(\omega-\omega_{\alpha}+\omega_{\alpha'})&&G_{\alpha\alpha',
\beta\beta'}(\mathbf{q},\omega)+ \\ \nonumber
&&P_{\alpha\alpha'}\sum_{\mu\nu}T_{\alpha\alpha',\nu\mu}(\mathbf{q})G_{\mu\nu,
\beta\beta'}(\mathbf{q},\omega)\\ \nonumber
&&=\frac{1}{2\pi}P_{\alpha\alpha'}\delta_{\alpha\beta'}\delta_{\beta\alpha'},
\end{eqnarray}
where $P_{\alpha\alpha'} = \langle L_{\alpha\alpha}\rangle - \langle
L_{\alpha'\alpha'}\rangle$ (we have suppressed the site indices $i$,
as we are using the homogeneous mean-field theory) and
$\omega_{\alpha}=V_{\alpha\alpha}+\sum_{\beta}T_{\alpha\alpha,\beta\beta}(q=0)\langle
L_{\beta\beta}\rangle$, and $\omega$ and $\mathbf{q}$ are,
respectively, the frequency and the wave vector. The propagator
$G^{ij}_{\alpha\alpha', \beta\beta'}(t,t')$ is the amplitude for the
$j^{th}$ site to flip between states $\beta$ and $\beta'$ at $t'$
given that the $i^{th}$ site has flipped between states $\alpha$ and
$\alpha'$ at $t$; and $\langle L_{\alpha\alpha}\rangle$ represents
the probability that the site is in the state $\alpha$ (we have
suppressed site indices). The poles of these Green functions give
the different branches of the excitation spectra in the RPA for
different phases in the BH model with two species of bosons.

\subsubsection{RPA Excitation Spectra for the Spin-1 BH Model}

To obtain the RPA excitation spectrum for the spin-1 BH model we use
$\hat{O} \equiv\sum_{\alpha\alpha'}\langle
i\alpha|\hat{O}|i\alpha'\rangle L_{\alpha\alpha'}^{i}$, so
Eq.(\ref{eq:bhspin1}) becomes (see the Appendix)
\begin{equation}
\label{eq:RPAESs1h}
{\cal{H}} = -\sum_{i\alpha\alpha'} V^{i}_{\alpha\alpha'}
L^{i}_{\alpha\alpha'}- \sum_{<i,j>,\alpha\alpha',\beta\beta'}
T^{ij}_{\alpha\alpha',\beta\beta'}
L^{i}_{\alpha\alpha'}L^{j}_{\beta\beta'},
\end{equation}
where
\begin{eqnarray}
\label{eq:RPAESs1v}
\nonumber
V^{i}_{\alpha\alpha'}&=&(\mu+\frac{U_0}{2})\langle
i\alpha|\hat{n}|i\alpha'\rangle -\frac{U_0}{2}\langle
i\alpha|\hat{n}^{2}|i\alpha'\rangle \\
&&-\frac{U_2}{2}\langle i\alpha|(\vec{F}^2 -
2\hat{n})|i\alpha'\rangle
\end{eqnarray}
and
\begin{eqnarray}
\label{eq:RPAESs1t}
\nonumber
T^{ij}_{\alpha\alpha',\beta\beta'} &=& \langle
i\alpha|\hat{a}^{\dagger}_{\sigma}|i\alpha'\rangle \langle
j\beta|\hat{a}_{\sigma}|j\beta'\rangle\\
&&+\langle
i\alpha|\hat{a}_{\sigma}|i\alpha'\rangle
\langle j\beta|\hat{a}^{\dagger}_{\sigma}|j\beta'\rangle,
\end{eqnarray}
where $\hat{n}\equiv\sum_\sigma\hat{n}_{\sigma}$,
$\hat{n}_{\sigma}\equiv a^{\dagger}_{\sigma}a_{\sigma}$,
$\vec{F}=\sum_{\sigma,\sigma '} a^{\dagger}_{\sigma}
\vec{F}_{\sigma ,\sigma '} a_{\sigma '}$, with $ \vec{F}_{\sigma
,\sigma '}$ standard spin-1 matrices, and $|i\alpha\rangle$ are the
mean-field eigenstates.

Again, by replacing averages of products of operators by products of
their averages, we obtain the Green functions, for this spin-1 BH
model, namely,
\begin{equation}
\label{eq:RPAESs1g}
G^{ij}_{\alpha\alpha', \beta\beta'}(t,t') = -i\theta(t-t')\langle[L^{i}_{\alpha\alpha'}(t);
 L^{j}_{\beta\beta'}(t')]\rangle
\end{equation}
and the RPA equation for its Fourier transforms:
 \begin{eqnarray}
\label{eq:RPAESs1w}
\nonumber
 &&(\omega-\omega_{\alpha}+\omega_{\alpha'})G_{\alpha\alpha',\beta\beta'}(\mathbf{q},\omega)\\
 \nonumber
  &&+P_{\alpha\alpha'}\sum_{\mu\nu}T_{\alpha\alpha',\nu\mu}(\mathbf{q})G_{\mu\nu,\beta\beta'}(\mathbf{q},\omega)
 =\frac{1}{2\pi}P_{\alpha\alpha'}\delta_{\alpha\beta'}\delta_{\beta\alpha',}\\
 &&
 \end{eqnarray}
where $P_{\alpha\alpha'} = \langle L_{\alpha\alpha}\rangle -
\langle L_{\alpha'\alpha'}\rangle$,  and
$\omega_{\alpha}=V_{\alpha\alpha}+\sum_{\beta}T_{\alpha\alpha,\beta\beta}(q=0)\langle
L_{\beta\beta}\rangle$.  The poles of these Greens functions give
the different branches of the excitation spectra in the RPA for
different phases in the spin-1 BH model.

\subsubsection{RPA Excitation Spectrum for the Extended Bose-Hubbard Model}

To obtain the RPA excitation spectrum for the EBH model, we write
Eq.(\ref{eq:ebh}) as (see the Appendix)
\begin{equation}
\label{eq:RPAESEBHh}
{\cal{H}} = -\sum_{i\alpha\alpha'} V^{i}_{\alpha\alpha'}
L^{i}_{\alpha\alpha'}- \sum_{<i,j>,\alpha\alpha',\beta\beta'}
T^{ij}_{\alpha\alpha',\beta\beta'}
L^{i}_{\alpha\alpha'}L^{j}_{\beta\beta'},
\end{equation}
where
\begin{equation}
\label{eq:RPAESEBHv}
V^{i}_{\alpha\alpha'}=(\mu+\frac{U}{2})\langle
i\alpha|\hat{n}|i\alpha'\rangle -\frac{U}{2}\langle
i\alpha|\hat{n}^{2}|i\alpha'\rangle
\end{equation}
and
\begin{eqnarray}
\label{eq:RPAESEBHt}
\nonumber
T^{ij}_{\alpha\alpha',\beta\beta'}&=&\langle i\alpha|\hat{a}^{\dagger}|i\alpha'\rangle \langle
j\beta|\hat{a}|j\beta'\rangle
+\langle i\alpha|\hat{a}|i\alpha'\rangle \langle
j\beta|\hat{a}^{\dagger}|j\beta'\rangle \\
&&-V \langle i\alpha|\hat{n}|i\alpha'\rangle \langle
j\beta|\hat{n}|j\beta'\rangle.
\end{eqnarray}

The Green functions, for this EBH model,
with the two sublattices ${\mathcal A}$ and ${\mathcal B}$, are
\begin{equation}
\label{eq:RPAESEBHg}
G^{ij}_{\alpha\alpha', \beta\beta'}(t,t') = -i\theta(t-t')\langle[L^{i}_{\alpha\alpha'}(t);
 L^{j}_{\beta\beta'}(t')]\rangle.
\end{equation}
In the RPA, the equations of motion for its
Fourier transforms are
 \begin{eqnarray}
\label{eq:RPAESEBHw1}
\nonumber
&&(\omega-\omega_{\alpha}+\omega_{\alpha'}){G}^{AA}_{\alpha\alpha', \beta\beta'}(\mathbf{q},\omega)\\
\nonumber
&&+P_{\alpha\alpha'}\sum_{\mu\nu}T_{\alpha\alpha',\nu\mu}(\mathbf{q})G^{BA}_{\mu\nu,\beta\beta'}
(\mathbf{q},\omega)=\frac{1}{2\pi} P_{\alpha\alpha'}\delta_{\alpha\beta'}\delta_{\beta\alpha'}\\
&&
\end{eqnarray}
\begin{eqnarray}
\label{eq:RPAESEBHw2}
\nonumber
&&(\omega-\omega_{\alpha}+\omega_{\alpha'}){G}^{BA}_{\alpha\alpha',
\beta\beta'}(\mathbf{q},\omega)\\
&&+P_{\alpha\alpha'}\sum_{\mu\nu}T_{\alpha\alpha',\nu\mu}(\mathbf{q})G^{AA}_{\mu\nu,
\beta\beta'}(\mathbf{q},\omega)= 0.
 \end{eqnarray}\\
The poles of these Green functions give the different branches of the
RPA excitation spectra for the different phases of the EBH model.

%
%

For each value of the frequency $\omega$ and the wave vector
$\mathbf{q}$, we obtain the eigenvalues of a $(2N_s-1)\times(2N_s-1)$ matrix that multiplies the Green function, where $N_s$ is the dimension of the number-state basis that we
use~\cite{sheshadri55,Konabe55,Menotti}. Thus, we obtain the RPA excitation spectra for these generalized BH models.

\section{Results}
\label{sec:results}

We present, in the next three subsections,  representative
results from our calculations of the RPA excitation spectra for
the phases of the BH model with two species of bosons, the spin-1 BH model, and
the EBH model.  In all our plots, we assume that our BH models
are defined on a simple square lattice, and the wave vector moves
along certain high-symmetry directions in the reciprocal lattice
of the simple square lattice (see, e.g.,~Ref. \cite{SCBZ}), in
particular, from $\Gamma$ $(q_x=0,q_y=0)$ to $X$ $(q_x=\pi,q_y=0)$, from $X$ to $T$ $(q_x=\pi,q_y=\pi)$, and then back
from $T$ to $\Gamma$.

\begin{figure}[htbp]
\centering \includegraphics[width=9cm,height=9cm]{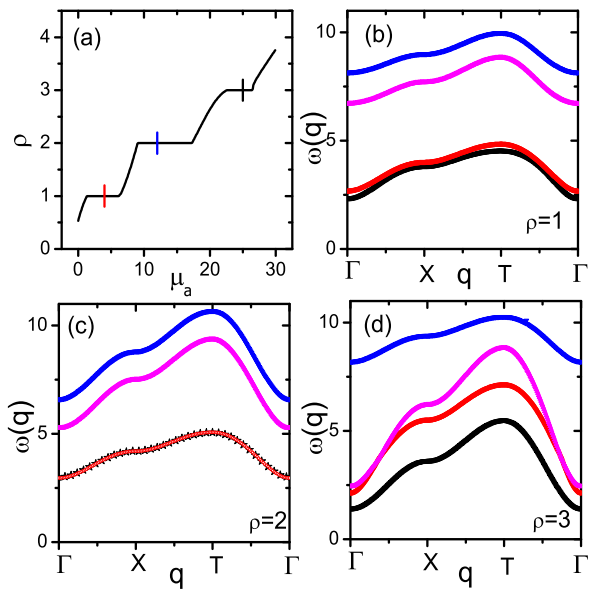}
\caption{(Color online) Plots of the excitation energy $\omega(q)$
versus the wave number $\textbf{q}$ (from $\Gamma$ to $X$, then $X$
to $T$, and finally $T$ to $\Gamma$, for a simple square
lattice~\cite{SCBZ}) for the BH model with two species of bosons.
These plots show representative excitation spectra in the MI phases.
(a) Total density $\rho$ versus $\mu_a$ for $U_{a}=11,~
U_{b}=0.9U_{a},~ \mu_{a}=\mu_{b}$ and $U_{ab}=0.6U_{a}$. The
vertical lines in the Mott plateaux represent the $\mu_a$ values for
which the excitation spectra were obtained for (b) $\rho=1$, (c)
$\rho=2$, and (d) $\rho=3$, with all other parameters the same in
all cases. For the MI phase with $\rho=1$, the excitation spectra
consist of one-hole and three-particle excitations. In contrast,
there are two- hole and particle excitations for $\rho = 2$ and $3$.
For the parameters chosen here, the lowest-energy excitation is
degenerate for $\rho=2$.} \label{fig:EKMI}
\end{figure}

\subsection{RPA Excitation Spectra for the Bose-Hubbard Model
with two species of bosons}

The results of our RPA calculations for excitation spectra in the
two-species BH model are shown by the representative plots in
Figs.\ref{fig:EKMI} - \ref{fig:EKSF}. First we consider the
excitation spectra in the MI phases,  with total density
of bosons $\rho=\rho_a+\rho_b=1$, $2$ and $3$, respectively, in
Figs.~\ref{fig:EKMI} (b)-(d). We also plot $\rho$ versus $\mu_a$ for
$U_a=11$, $U_b=0.9U_a$, $U_{ab}=0.6U_{a}$  and $\mu_b=\mu_a$ in Fig.
\ref{fig:EKMI}(a) and mark, with vertical lines, the $\mu_a$ values
for which the spectra was obtained in Figs.~\ref{fig:EKMI} (b)-(d).
The excitation spectra in the MI phase with $\rho=1$, given in Fig.
\ref{fig:EKMI}(b), show four excitations, one of which is a
hole excitation and the rest are particle excitations (see
Eq.~(\ref{eq:2bhb6}) from Appendix for details). However, for
$\rho=2$ and $\rho=3$, the spectra, plotted, respectively, in Figs.
\ref{fig:EKMI} (c) and (d), show four branches;
these correspond to two-hole and two-particle excitations,
which are the solutions of Eqs.~{\ref{eq:2bhb4} and \ref{eq:2bhb5}.
For $\rho=2$, two excitations are almost degenerate for the
parameters we have considered. In all these cases we also obtain
dispersionless modes, which are independent of ${\bf q}$ modes; we
described these in the Appendix so we do not show them here.

\begin{figure}[htbp]
\centering \includegraphics[width=9cm,height=6cm]{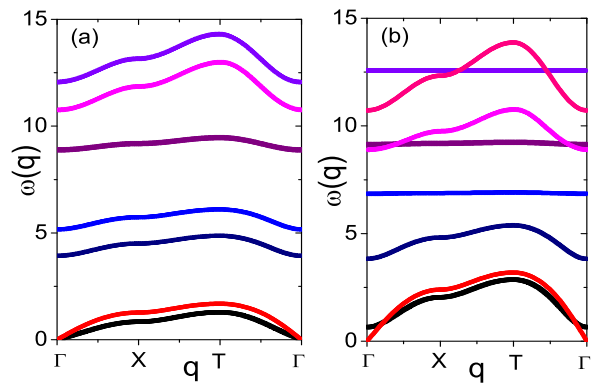}
\caption{(Color online) Plots of excitation spectra (a) when both
types of bosons are in the SF phase and (b) when only one type is in
the SF and the other type of boson is in MI phase. }
\label{fig:2EKMI}
\end{figure}
\begin{figure}[htbp]
\centering \includegraphics[width=9cm,height=9cm]{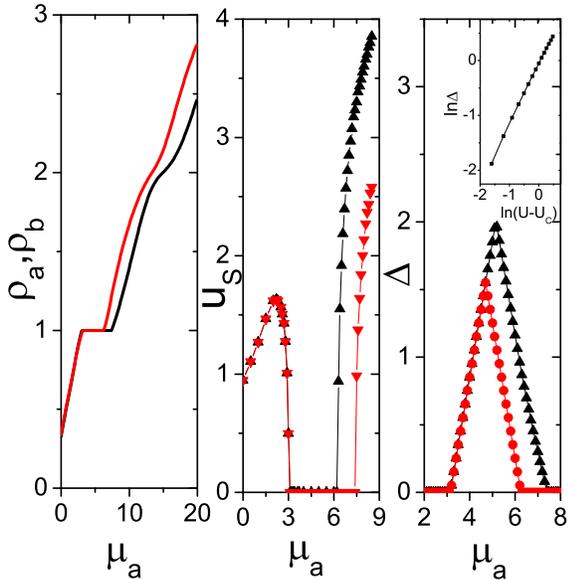}
\caption{(Color online) (a) Plots of $\rho_a$ (red line) and
$\rho_b$ (black line), for bosons of types $a$ and $b$, versus
$\mu_{a}$ for $U_{a}=8,~ U_{b}=0.9U_{a},~ \mu_{a}=\mu_{b}$, and
$U_{ab}=0.2U_{a}$; note the plateaux at integer values of $\rho_{a}$
and $\rho_{b}$. (b) The speed of sound $u_s$ versus $\rho_a$, for
$U_a = 8,~ U_b = 0.9U_a,~ \rho_a =\rho_b$, and $U_{ab} = 0.2U$;
$u_s$ is zero in the MI phase and finite in the SF phase. (c) The
gap $\Delta$ versus $\mu_a$, for $U_{a}=8,~ U_{b}=0.9U_{a},~
\mu_{a}=\mu_{b}$ and $U_{ab}=0.2U_{a}$; the inset shows the
power-law behavior of the gap in the vicinity of the transition at
$U_{c}=3.1$.} \label{fig:3EKMI}
\end{figure}
\begin{figure}[htbp]
\centering \includegraphics[width=9cm,height=9cm]{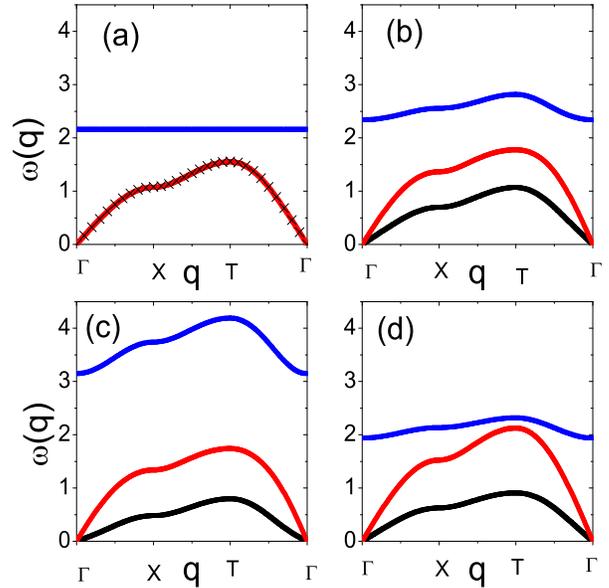}
\caption{(Color online) Plots of the excitation energy $\omega(\bf{q})$ versus
the wave number $\textbf{q}$ for
the BH model with two species of bosons, for $\rho = 0.75,$
$U_{a}=9,~ U_{b}=0.9U_{a}$, and (a) $U_{ab}=0.0$, (b)
$U_{ab}=0.2U_{a}$, (c) $U_{ab}=0.4U_{a}$ and (d) $U_{ab}=-0.2U_{a}$.
} \label{fig:EKSF}
\end{figure}

The excitation spectra for the case where both $a$-type and $b$-type bosons are
in the SF phase are given in the Fig. \ref{fig:2EKMI}(a) which show two gapless
excitations; linear in wave number near the point $\Gamma$ and excitations
which have finite gap at the point $\Gamma$. Similarly, the excitation spectra
for the case where only  $b$-type bosons are in the SF phase and $a$-type
bosons are in the MI phase are given in Fig. \ref{fig:2EKMI}(b); we see only
one gapless excitation corresponding to the SF phase of $b$-type bosons; and
all the other branches of the excitation spectra have finite gaps at the point
$\Gamma$. We also show some of the dispersionless spectra.

In Fig. \ref{fig:3EKMI} (a), we present plots versus $\mu_a = \mu_b$
of $\rho_a$ (red curve) and $\rho_b$ (black curve), for bosons of
types $a$ and $b$, for $U_{a}=8,~ U_{b}=0.9U_{a},~ \mu_{a}=\mu_{b}$
and $U_{ab}=0.2U_{a}$, with plateaux at $\rho_a = 1$ and/or $\rho_b
= 1$. In a gapless, SF phase, the speed of sound $u_s$
follows~\cite{sheshadri55} from the slope of the excitation spectrum
near the point $\Gamma$.  In Fig. \ref{fig:3EKMI} (b), we show how
the speed of sound $u_s$ behaves, for components $a$ (red dashed
line) and $b$ (black dashed line with triangles), as functions of
$\mu_a$, for the same parameters as in Fig. \ref{fig:3EKMI} (a); as
we expect, $u_s$ is zero in the MI phases and positive in the SF
phases.

From the excitation spectra we can obtain the gap $\Delta$ at the
point $\Gamma$. In Fig. \ref{fig:3EKMI} (c), we show plots of
$\Delta$ versus $\mu_a$ for the same parameters as in Figs.
\ref{fig:3EKMI} (a) and (b); as we expect, $\Delta$ is zero in the
SF phases (for bosons of types $a$ and $b$) and finite in the MI
phases. In the inset of Fig. \ref{fig:3EKMI} (c) we shows that
$\Delta$ approaches zero as $(U-U_c)^{1/2}$, where $U_c$ is the
critical value of $U$ at which the MI-SF transitions occur at these
parameter values.

We now consider the first few low-energy excitation spectra in the SF phase; we
hold $\rho=0.75$, but vary the $U_{ab}$ in Figs.  \ref{fig:EKSF} (a)-(c). In
all these figures $U_a=9$, $U_b=0.9U_a$ and $\mu_a=\mu_b$. In Fig.
\ref{fig:EKSF} (a), $U_{ab}/U_a=0$, we see that the gapless mode is almost
degenerate; and the next branch in the excitation spectrum is dispersionless;
we do not show high-energy spectra here. As we increase $U_{ab}/U_{a}$, Fig.
\ref{fig:EKSF} (b) for  $U_{ab}/U_a=0.2$ and Fig. \ref{fig:EKSF} (c) for
$U_{ab}/U_a=0.4$, the degeneracy of the gapless excitation spectrum is lifted
and the spectra move away from each other. The non-dispersive mode now shows
dispersion and moves to higher energies. In Fig.  \ref{fig:EKSF} (d), we
illustrate the excitation spectra for $U_{ab}/U_{a} < 0$; we see again that the
gapless modes move away from each other; and we observe that the excitation
frequencies of other modes are different compared to their values for
$U_{ab}/U_{a}
> 0$. Our results are qualitatively similar to those of the
Gutzwiller-based, excitation-spectrum study of Ref.~\cite{ozaki0955}, which
presents $k_x=k_y$ scans through the reciprocal lattice of a two-dimensional
square lattice.
\begin{figure}[htbp]
\centering \includegraphics[width=9cm,height=7cm]{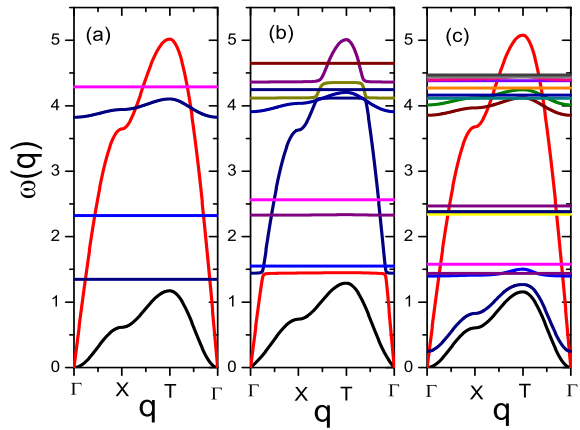}
\caption{(Color online) Plots of the excitation energy $\omega$
versus the wave number $\textbf{q}$  for the spin-1 BH model: (a)
superfluid ($U_{0}=2,~\mu=5$ and $~U_{2}/U_{0}=0$), (b)
polar-superfluid ($U_{0}=2,~\mu=5$ and $~U_{2}/U_{0}=0.03$) and (c)
ferromagnetic-superfluid ($U_{0}=2,~\mu=5$ and $U_{2}/U_{0}=-0.03$)
phases. In (a) and (b) the lowest modes are two gapless, degenerate
modes, shown in black, whose energy approaches zero quadratically at
small wave numbers $\textbf{q}$ (near $\Gamma$). In (c) we see
clearly the splitting of the degenerate modes, which now have a gap
and then a quadratic dependence on $\textbf{q}$ near $\Gamma$. In
all three cases,  there is another mode, shown in red, whose energy
goes to zero linearly at small $\textbf{q}$. All the higher
excitations have finite gaps at $\Gamma$ and are highly degenerate.
The degeneracies of these modes are partially lifted in the case of
polar and ferromagnetic SF phases.} \label{fig:Ekspin1_1}
\end{figure}

\subsection{RPA Excitation Spectra for the Spin-1 Bose-Hubbard Model}

We now give representative plots of our RPA excitation spectra for the spin-1
BH model, in  Figs. \ref{fig:Ekspin1_1} (a)-(c) for $~U_{2}/U_{0}=0$,
$~U_{2}/U_{0}=0.03$, and $U_{2}/U_{0}=-0.03$, which yield, respectively, a pure
SF with no spin interactions, a polar SF, and a ferromagnetic SF for the
parameter values given in the figure caption~\cite{rvpspin155}. The excitation
spectra are qualitatively different in these three cases.  In the SF (Fig.
\ref{fig:Ekspin1_1} (a)) phase there are two gapless, degenerate modes, shown
in black, whose frequencies approach zero quadratically at small wave numbers
$\textbf{q}$ (near $\Gamma$); there is another mode, shown in red, whose energy
goes to zero linearly at small $\textbf{q}$. All the other excitations spectra
have finite gap at $\Gamma$; some of these spectra are degenerate and some are
${\bf q}$ independent; e.g., the spectra shown in green, blue, and maroon are
${\bf q}$ independent and have degeneracies 3, 4, and 8, respectively. The
branch shown in navy blue is weakly $\textbf{q}$ dependent and is doubly
degenerate. These degeneracies are partially lifted when $U_2$ is finite. In
the polar SF (Fig.  \ref{fig:Ekspin1_1} (b)), the lowest modes, shown in black,
are doubly degenerate and have a quadratic dependence on $\textbf{q}$ as in the
case of a pure SF. However, the next branch, which displays a linear dependence
on $\textbf{q}$ near the $\Gamma$ point, as in the case of a pure SF, becomes
approximately dispersionless at higher value of $\textbf{q}$, and avoids level
crossing with the branch shown in green (these acquire a ${\bf q}$
dependence). Note that this spectrum, dispersionless and with a triple
degeneracy in the pure SF case, has its degeneracy lifted partially: the
spectrum shown in navy blue is non-degenerate and the one shown in blue is
doubly degenerate; we obtain similar behaviors in higher-energy excitation
spectra. In the ferromagnetic SF (Fig. \ref{fig:Ekspin1_1} (c)), we see clearly
the splitting of the degenerate modes, which now have a gap and then a
quadratic dependence on $\textbf{q}$ near $\Gamma$; the red curve indicates the
gapless mode, which approaches $\Gamma$ linearly. The degeneracies of higher
modes are also lifted in this case. The qualitative features of these spectra
agree with those found for the various SF phases in the spin-1,
Gross-Pitaevskii study of Ref.~\cite{tlho55}.

\begin{figure}[htbp]
\centering \includegraphics[width=9cm,height=9cm]{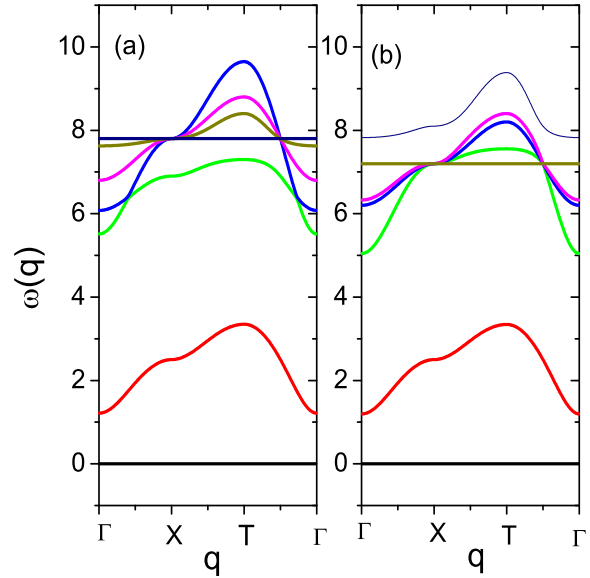}
\caption{(Color online) Plots of the excitation frequency
$\omega({\bf q})$ versus the wave number $\textbf{q}$ for the spin-1
BH model in the MI1 phase with $\rho = 1$: (a) $U_0 = 10$, $\mu=2.5$
and $U_2/U_0=0.03$ and (b) $U_0 = 10$, $\mu=2.5$ and $U_2/U_0 =
-0.03$. The MI1 ground state is degenerate; and these excited states
do not couple to the ground state through particle or hole
excitations; this gives the the dispersionless mode at
$\omega(\bf{q}) = 0$. The mode shown in red is a hole excitation;
all the higher-energy modes shown are particle excitations. }
\label{fig:Ekspin1_2}
\end{figure}
\begin{figure}[htbp]
\centering \includegraphics[width=9cm,height=9cm]{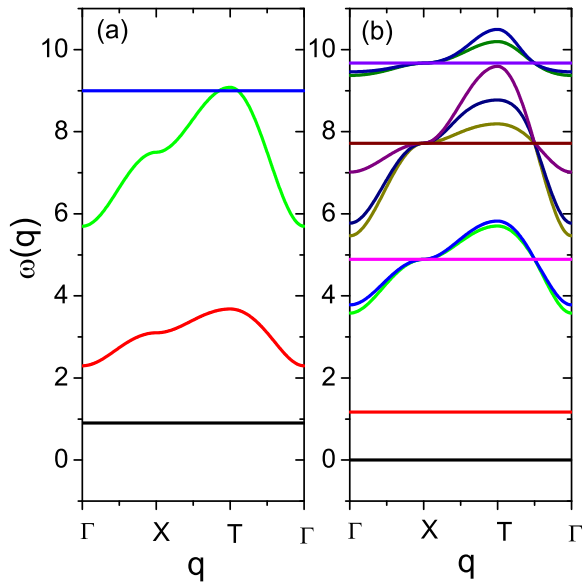}
\caption{(Color online) Plots of the excitation energy
$\omega(\bf{q})$ versus the wave number $\textbf{q}$  for the spin-1
BH model with $\rho=2$: (a) $U_{0}=10,~\mu=12.5$ and
$~U_{2}/U_{0}=0.03$ and (b) $U_{0}=13,~\mu=17.5$ and
$U_{2}/U_{0}=-0.03$. } \label{fig:Ekspin1_3}
\end{figure}

\begin{figure}[htbp]
\centering \includegraphics[width=8cm,height=6cm]{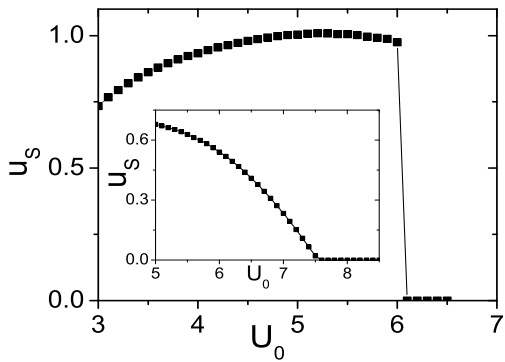}
\caption{The sound speed $u_s$ versus $U_{0}$ in the vicinity of the
polar-SF-MI2 transition for $\mu=10$ and $U_{2}/U_{0}=0.03$, with
$(\rho=2)$; $u_s$ jumps to zero at this first-order transition; by
contrast, (see inset for $~\mu=5$ and $U_{2}/U_{0}=0.03$) $u_s$ goes
to zero continuously at the SF-MI1 transition.}
\label{fig:Ekspin1_4}
\end{figure}
We present excitation spectra in the MI1 and MI2 phases in
Figs.~\ref{fig:Ekspin1_2} and ~\ref{fig:Ekspin1_3}, respectively.  Here, MI1
and MI2 denote, respectively, Mott phases with one and two bosons per site.
For MI1, we obtain qualitatively similar excitation spectra for $U_2/U_0 > 0$
(Figs.~\ref{fig:Ekspin1_2}(a)) and $U_2/U_0 < 0$
(Figs.~\ref{fig:Ekspin1_2}(b)). For $U_2/U_0 > 0$, the ground state for MI1 is
triply degenerate (Appendix); these degenerate states do not couple to each
other through the creation of a particle or hole. Thus, the lowest excitation,
$\omega({\bf q}) = 0$, represented in black in Fig.~\ref{fig:Ekspin1_2}(a), is
doubly degenerate and dispersionless. The non-degenerate hole excitation is
shown in red. The excitations, above the hole excitation, are six-particle
excitations, two of which are dispersionless and degenerate. We find similar
excitation spectra for  $U_2 /U_0 < 0$ (Figs.~\ref{fig:Ekspin1_2}(b)).

For MI2, we obtain qualitatively different excitation spectra for $U_2/U_0 > 0$
(Figs.~\ref{fig:Ekspin1_3}(a)) and $U_2/U_0 < 0$
(Figs.~\ref{fig:Ekspin1_3}(b)). For $U_2/U_0 > 0$, the MI2 ground state is
non-degenerate (Appendix) because of the formation of a  singlet, so there is
no dispersionless mode with $\omega(\bf{q})=0$.  The first excited state of our
MFT hamiltonian (Eq.~\ref{eq:mfhspin1}) has five-fold degeneracy; and the
ground state does not couple to these states via particle or hole excitations.
This gives us the dispersionless mode (black line). The dispersive mode with
the lowest energy comprises hole excitations (red curves); this mode is  triply
degenerate. The particle excitations are also dispersive (green curve,
degeneracy 3), except for the one shown in blue (degeneracy 6).  Higher-energy
excitations are multi-particle-hole excitations and are dispersionless; we do
not show these here.

In Fig.~\ref{fig:Ekspin1_4} we plot the sound speed $u_s$ versus $U_{0}$ in the
vicinity of the polar-SF-MI2 transition for $\mu=10$ and $U_{2}/U_{0}=0.03$,
with $(\rho=2)$; $u_s$ jumps to zero at this first-order transition; by
contrast, (see inset for $~\mu=5$ and $U_{2}/U_{0}=0.03$) $u_s$ goes to zero
continuously at the phase-SF-MI1 transition. The natures of these transitions
are consistent with the predictions of the mean-field theory~\cite{rvpspin155}
on which we base our RPA study.

We do not investigate magnetic ordering and magnetic excitations in the MI
phase of the spin-1 model because, as has been noted earlier, our mean-field
theory does not allow for any magnetic structure in the MI phase in this
model~\cite{rvpspin155}.

\subsection{RPA Excitation Spectra for the Extended Bose-Hubbard model}

We now show representative RPA excitation spectra in
Figs. \ref{fig:Ekebh} (a), (b), (c), (d), and (e) for SF, DW
$3/2$ (see Ref.~\cite{JMK12}), SS , MI1, and DW $1/2$ phases for $V=0.6U$ and
(a) $U=6,~\mu=11.3$, (b) $U=10,~\mu=19$, (c) $U=8.3,~\mu=15.4$,
(d) $U=12,~\mu=10$, and (e) $U=12,~\mu=5$. By the symbol DW $n/m$, we
mean a density-wave phase with $n$ bosons per site and $m$ sites per
unit cell. From these excitation spectra we obtain $u_s$, whose
dependence on $U$ is shown in Fig. \ref{fig:Ekebh} (f), as the
system moves from the SF to the SS phase and then to the DW $3/2$
phase, for $V=0.6U$, and $\rho_{total}=3$; the sharp drop in $u_s$ at
the SS-DW $3/2$ boundary indicates a first-order transition; the
inset shows that $u_s\sim\sqrt{U}$ in the small-$U$ regime in the SF phase.

The RPA excitation spectra have a gap at the  $\Gamma$ point when the system is
in the DW 3/2 phase and this gap $\Delta$ goes to zero as the system approaches
the SS phase. In Fig. \ref{fig:Ekebh2} $(a)$ we plot $\Delta$ versus $U$ as the
system goes from the DW to the SS phase.  Figure \ref{fig:Ekebh2} $(b)$ show
analogous plots for the SF-MI transition. The behaviors of this gap are what we
expect on general grounds, i.e., it jumps discontinuosly to zero at the DW
$3/2$ to SS transition, but it goes to zero continuously, as $\Delta \sim
(U-U_c)^{1/2}$, at the MI1 to SF transition at the critical value $U_c$. This
exponent of $1/2$ is a mean-field exponent, which should be modified by
fluctuations (this MI1-SF transition lies in the $d$-dimensional XY
universality class).

\begin{figure}[htbp]
\centering \includegraphics[width=9cm,height=9cm]{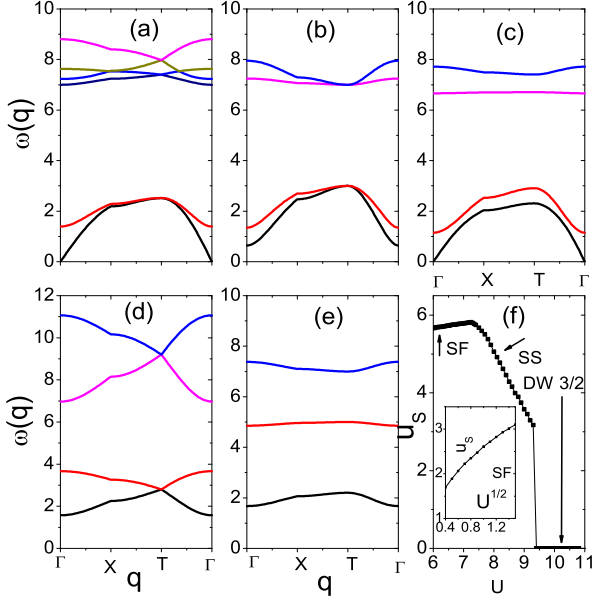}
\caption{(Color online) Plots of the excitation frequency
$\omega(\bf{q})$ versus the wave number $\textbf{q}$  for the EBH
model in (a) $U=6,~V=0.6U,~\mu=11.3$, and $\rho_{total}=3$, (b)
$U=10,~V=0.6U,~\mu=19,~\rho_{total}=3$, (c)
$U=8.3,~V=0.6U,~\mu=15.4,~\rho_{total}=3$, (d)
$U=12,~V=0.6U,~\mu=10,~\rho_{total}=2$, and (e)
$U=12,~V=0.6U,~\mu=5,~\rho_{total}=1$. We show the excitation
spectra in the SF, DW3/2, SS, MI1, and DW1/2 phases, respectively.
In (f) $V=0.6U,~\rho_{total}=3$; and we show the dependence of $u_s$
on $U$ as the system goes from the SF to the SS and then to the
DW3/2 phase; the last transition is clearly first order; the inset
shows the $u_s\sim\sqrt{U}$ behavior at low $U$ in the SF phase.}
\label{fig:Ekebh}
\end{figure}


\begin{figure}[htbp]
\centering \includegraphics[width=9cm,height=9cm]{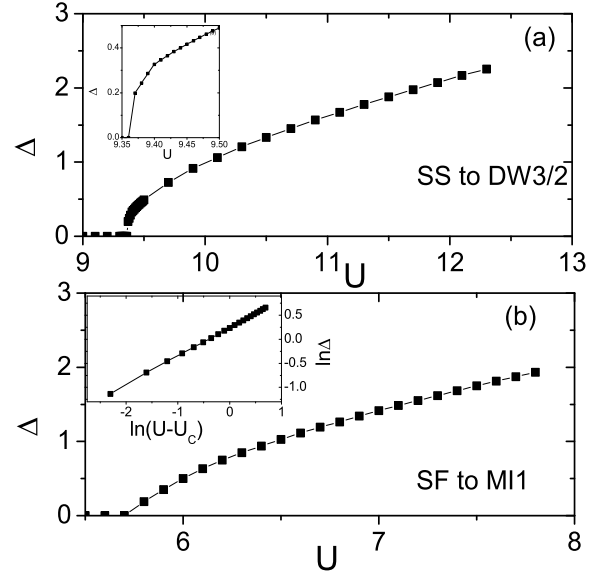}
\caption{Plots of the gap $\Delta$ in the RPA excitation spectrum as
a function of $U$ (a) near the SS-DW3/2 transition and  $(b)$ near
the SF-MI transition; the inset shows the power-law behavior of the
gap in the vicinity of the transition at $U_{c}=5.7$.}
\label{fig:Ekebh2}
\end{figure}
\section{Conclusions}\label{conclusions}

We have obtained representative excitation spectra in all phases of three
generalized Bose-Hubbard (BH) models, namely, (1) a two-species
generalization of the spinless BH model, (2) a spin-1 BH model,
and (3) the extended Bose-Hubbard model (EBH) for spinless
interacting bosons of one species. Our study uses the random
phase approximation (RPA), which it develops, in a unified way,
by starting from mean-field theories of the type that we have
discussed in Ref.~\cite{rvpspin155}. Thus, it generalizes the RPA studies
initiated in Ref.~\cite{sheshadri55} for the simple BH model and
continued, e.g., in Refs.~\cite{Konabe55,Menotti,iskin09}.

Our study yields a variety of interesting results that we have
described in detail in the previous Section. In particular, our
RPA excitation spectra show clear gaps in MI phases, in all the
models above, in the DW phases in the EBH model, and gapless
spectra in all SF phases and the SS phase in the EBH model. From
these spectra we have obtained the dependence of (a) gaps
$\Delta$ and (b) the sound velocity $u_s$ on the parameters of
these models. We have also investigated the behaviors
of $\Delta$ and $u_s$ as these systems go through phase
transitions. We find that, at the polar SF-MI transitions in the
spin-1 BH model, $u_s$ goes to zero continuously
(discontinuously) for MI phases with an odd (even) number of
bosons per site; this is consistent with the natures of these
transition (continuous or discontinuous) in the mean-field theory
for the spin-1 BH model~\cite{rvpspin155}.

In the SF phases of these models, our excitation spectra agree
qualitatively, at weak couplings, with those that can be obtained
from Gross-Pitaevskii-type models. For example, our RPA
excitation spectra are qualitatively similar to those obtained
for ferromagnetic and polar superfluids in~\cite{tlho55}, which
uses a spin-1 generalization of the Gross-Pitaevskii equation.

Excitation spectra, at the level of the RPA, can also be obtained by starting
with a Gutzwiller-type mean-field theory. Such studies, carried out, e.g., for
two-boson BH models, the spin-1 BH models, and the EBH model, should be
formally equivalent to our RPA study; however, we are not aware of any study
that has shown this in complete detail. We do not know of one, unified
treatment of Gutzwiller-type RPA, like our unified RPA study for these three
models, so a direct comparison of our results with these Gutzwiller-type
studies is not completely straightforward.  Reference~\cite{Hou} has evaluated
the excitation spectrum of spin-1 bosonic atoms in a Mott-insulator phase. Hou,
\textit{et al.} have investigated the quantum phase transition of spin-2 cold
bosons in an optical lattice with and without an external magnetic
field,~\cite{Hou2}; they also obtain Gutzwiller-type RPA excitation spectra.

Previous studies have analyzed excitation spectra in the two-species BH model.
In the MI phase, the lowest two branches of the excitation spectrum have gaps;
and they correspond to the particle- and hole-excitation
modes~\cite{Elstner55,Oosten55,Altman55,Konabe55}. In the SF phase, the
excitation spectrum has one gapless mode and one mode with a
gap~\cite{E.Altman55,Huber55,Krutitsky55,Podolsky55,Pollet55}; the gapless mode
arises because of oscillations of the phase of the SF order parameter (this is
the Bogoliubov mode); the lowest mode with the gap arises because of the
amplitude mode in the vicinity of the SF-MI transition at integer fillings;
this amplitude mode becomes gapless at the critical point. Our study explores
the dependence of the excitation spectra on model parameters in far greater
detail than earlier studies.

Furthermore, the RPA method can be used to obtain the complete $\omega$ and
$\bf{q}$ dependence of the Green function as noted, e.g., in
Refs.~\cite{Konabe55,Menotti} for the simple, spinless BH model. From this,
various properties can be calculated, e.g., the momentum distribution function;
the latter has been obtained in the RPA in~\cite{iskin09} for the EBH model. We
give a brief description of such Green-function studies in the Appendix and
their generalzations for the three models we consider here.

Excitation spectra have been measured experimentally by
Bragg-spectroscopy~\cite{Stenger55, Steinhauer55, Ernst55, Clement55,
Bissbort55, Miyake55, Fabbri55} and
lattice-amplitude-modulation~\cite{Stoerle55, Schori55, Endres55} methods; and
these measurements have been used to characterize SF and MI phases and the
transition between them. We hope our work will lead to experimental
measurements of the spectra of elementary excitations in the different phases
in the physical realizations of the generalized BH models we have discussed
above.  Momentum-distribution functions can be calculated, at the RPA level,
from the Green functions that we have obtained. The calculation is especially
simple in a Mott phase, as illustrated, e.g., in Ref.~\cite{iskin09}. Such
momentum-distribution functions can be compared with those that are measured
experimentally as discussed, e.g., in Refs.~\cite{rmp55,iskin09}.  The studies
of Refs. \cite{Menotti,iskin09} also mention some of the limitations of the RPA
in the context of the BH model; e.g., Ref. \cite{Menotti} notes that this type
of of RPA, when applied to the simple BH model, yields a violation of the total
density sum rule.

Excitation spectra can also be measured by Quantum Monte Carlo (QMC) simulation
as shown for the simple BH model in Ref.~\cite{mc55} and a continuation model
in Ref.~\cite{Saccani}. We hope our study will lead to QMC studies of the
excitation spectral of the generalized BH models we have discussed here.

\vspace{1cm}

\section{Acknowledgments}

We thank DST, UGC, CSIR (India)(No. 03(1306)/14/EMR-II) for support and K. Sheshadri for useful
discussions. JMK and RVP thank, respectively, the University of Goa and the
Indian Institute of Science, for hospitality during the period in which this
paper was being written.

\appendix
\section{The Random-Phase Approximation.}
We illustrate below how we obtain the RPA equations for the Green functions for our Bose-Hubbard (BH) models of Eqs. (1), (2), and (3), respectively.

The RPA equation of motion, for any of these BH models use the operators
$L_{\alpha\alpha'}^{i}=|i\alpha\rangle \langle i\alpha'|$, where $|i\alpha\rangle$ is the eigenvector for the mean-field Hamiltonian
for site $i$. Any single-site operator can be expressed as
$\hat{O} \equiv\sum_{i\alpha\alpha'}\langle i\alpha|\hat{O}|i\alpha'\rangle L_{\alpha\alpha'}^{i}$,
 so the Hamiltonian for these BH models can be written as

\begin{eqnarray}
\label{eq:RPAEShm}
\nonumber
{\cal{H}}^{m} = &&-\sum_{i\alpha\alpha'} V^{i,m}_{\alpha\alpha'} L^{i,m}_{\alpha\alpha'}
-\sum_{ij,\alpha\alpha',\beta\beta'} T^{ij,m}_{\alpha\alpha',\beta\beta'} L^{i,m}_{\alpha\alpha'}L^{j,m}_{\beta\beta'},\\
&&
\end{eqnarray}
where the superscript $m=1$, is for two species of bosons, $m=2$ is for the spin-1 BH model, and $m=3$ is for the EBH model.

\begin{eqnarray}
\label{eq:RPAESvm1}
V^{i,1}_{\alpha\alpha'}=&&(\mu_{a}+\frac{U_a}{2})\langle
i\alpha|\hat{n}_a|i\alpha'\rangle -\frac{U_a}{2}\langle
i\alpha|\hat{n}_a^{2}|i\alpha'\rangle\nonumber \\
&&+(\mu_{b}+\frac{U_b}{2})\langle i\alpha|\hat{n}_b|i\alpha'\rangle
-\frac{U_b}{2}\langle i\alpha|\hat{n}_b^{2}|i\alpha'\rangle\nonumber\\
&&-U_{ab}\langle i\alpha|\hat{n}_a\hat{n}_b|i\alpha'\rangle,
\end{eqnarray}

\begin{eqnarray}
\label{eq:RPAESvm2}
 V^{i,2}_{\alpha\alpha'}&=&(\mu+\frac{U_0}{2})\langle
i\alpha|\hat{n}_i|i\alpha'\rangle -\frac{U_0}{2}\langle
i\alpha|\hat{n}_i^{2}|i\alpha'\rangle\nonumber \\
&&- \frac{U_2}{2}\langle i\alpha|(\vec{F}^2_i -
2\hat{n}_i)|i\alpha'\rangle,
\end{eqnarray}

\begin{equation}
\label{eq:RPAESvm3}
V^{i,3}_{\alpha\alpha'}=(\mu+\frac{U}{2})\langle
i\alpha|\hat{n}_i|i\alpha'\rangle -\frac{U}{2}\langle
i\alpha|\hat{n}_i^{2}|i\alpha'\rangle,
\end{equation}
and
\begin{widetext}
\begin{equation}
\label{eq:RPAEStm1}
T^{ij,1}_{\alpha\alpha',\beta\beta'} = \langle i\alpha|\hat{a}_i^{\dagger}|i\alpha'\rangle
\langle j\beta|\hat{a}_j|j\beta'\rangle+\langle i\alpha|\hat{a}_i|i\alpha'\rangle
\langle j\beta|\hat{a}_j^{\dagger}|j\beta'\rangle
+\langle i\alpha|\hat{b}_i^{\dagger}|i\alpha'\rangle
\langle j\beta|\hat{b}_j|j\beta'\rangle+\langle i\alpha|\hat{b}_i|i\alpha'\rangle
\langle j\beta|\hat{b}_j^{\dagger}|j\beta'\rangle,
\end{equation}

\begin{equation}
T^{ij,2}_{\alpha\alpha',\beta\beta'} =\sum_{\sigma}\{ \langle i\alpha|\hat{a}_{i,\sigma}^{\dagger}|i\alpha'\rangle \langle j\beta|\hat{a}_{j,\sigma}|j\beta'\rangle
+\langle i\alpha|\hat{a}_{i,\sigma}|i\alpha'\rangle
\langle j\beta|\hat{a}_{j,\sigma}^{\dagger}|j\beta'\rangle \},
\end{equation}

\begin{equation}
T^{ij,3}_{\alpha\alpha',\beta\beta'} = \langle i\alpha|\hat{a}_i^{\dagger}|i\alpha'\rangle
\langle j\beta|\hat{a}_j|j\beta'\rangle+\langle i\alpha|\hat{a}_i|i\alpha'\rangle
\langle j\beta|\hat{a}_j^{\dagger}|j\beta'\rangle - V \langle i\alpha|\hat{n}_i|i\alpha'\rangle \langle j\beta|\hat{n}_j|j\beta'\rangle.
\end{equation}
\end{widetext}

The propagator $G^{ij,m}_{\alpha\alpha', \beta\beta'}(t,t')$, is the amplitude
for the $j^{th}$ site to flip between the states $\beta$ and $\beta'$ at
$t'$, given that the $i^{th}$ site has flipped between states $\alpha$ and
$\alpha'$ at $t$. We have
\begin{eqnarray}
\label{eq:RPAESgm}
\nonumber
G^{ij,m}_{\alpha\alpha', \beta\beta'}(t,t') &=& -i\theta(t-t')\langle[L^{i,m}_{\alpha\alpha'}(t),
 L^{j,m}_{\beta\beta'}(t')]\rangle \\
 &=&\langle\langle L^{i,m}_{\alpha\alpha'}(t)|L^{j,m}_{\beta\beta'}(t')\rangle\rangle.
\end{eqnarray}
The equation of motion for this Green function, with $t'=0$, is calculated
as follows:

\begin{eqnarray}
\label{eq:RPAESdgm}
\nonumber
\frac{d}{dt}G^{ij,m}_{\alpha\alpha', \beta\beta'}(t) &=& -i\frac{d\theta(t)}{dt}\langle\left[L^{i,m}_{\alpha\alpha'}(t), L^{j,m}_{\beta\beta'}(0)\right]\rangle \\
&&- i\theta(t)\frac{d}{dt}\langle\left[L^{i,m}_{\alpha\alpha'}(t), L^{j,m}_{\beta\beta'}(0)\right]\rangle,
\end{eqnarray}
where
\begin{eqnarray}
\label{eq:RPAESdlm}
\nonumber
&&\frac{d}{dt}L^{i,m}_{\alpha\alpha'}(t) = \frac{1}{i\hbar}\left[ {\cal{H}}^m, L^{i,m}_{\alpha\alpha'}(t)\right]
\equiv i\left[L^{i,m}_{\alpha\alpha'}(t), {\cal{H}}^m\right]\\
&&
\end{eqnarray}
we have set $\hbar = 1$, and we obtain the second term on the right-hand side of Eq.~(\ref{eq:RPAESdgm})

\begin{eqnarray}
\label{eq:RPAESdlm2}
-i\frac{d}{dt}&&\left[L^{i,m}_{\alpha\alpha'}(t), L^{j,m}_{\beta\beta'}(0)\right]=\left[L^{i,m}_{\alpha\alpha'}(t),{\cal{H}}^m\right]L^{j,m}_{\beta\beta'}(0) \nonumber \\
&&-L^{j,m}_{\beta\beta'}(0)\left[L^{i,m}_{\alpha\alpha'}(t),{\cal{H}}^m\right]
\end{eqnarray}

By using Eq.~(\ref{eq:RPAEShm}) in the above equation, substituting it in Eq.~(\ref{eq:RPAESdgm}) and carrying out the
RPA, where we replace thermal averages of products of operators by the products of their thermal averages, to obtain

\begin{widetext}
\begin{equation}
\label{eq:RPAESdgm2}
\frac{d}{dt}G^{ij,m}_{\alpha\alpha', \beta\beta'}(t)=-i\delta(t)\delta_{\alpha'\beta}\delta_{\alpha\beta'}
{P}^{m}_{\alpha\alpha'}+i({\omega}^{m}_{\alpha'}-{\omega}^{m}_{\alpha})G^{ij,m}_{\alpha\alpha', \beta\beta'}(t)
+i{P}^{m}_{\alpha\alpha'}\sum_l\sum_{\mu\nu}T^{il,m}_{\alpha\alpha',\nu\mu}G^{lj,m}_{\mu\nu, \beta\beta'}(t).
\end{equation}

Fourier transformation over space and time for $m=1$ and $m=2$ now yields
\begin{equation}
\label{eq:RPAESwm}
 (\omega-{\omega}^{m}_{\alpha}+{\omega}^{m}_{\alpha'})G^{m}_{\alpha\alpha', \beta\beta'}(\mathbf{q},\omega)
+{P}^{m}_{\alpha\alpha'}\sum_{\mu\nu}T^m_{\alpha\alpha',\nu\mu}(\mathbf{q})G^{m}_{\mu\nu, \beta\beta'}
\mathbf{q},\omega)=\frac{1}{2\pi}{P}^{m}_{\alpha\alpha'}\delta_{\alpha\beta'}\delta_{\beta\alpha'};
\end{equation}

and, for $m=3$ with a bipartite hypercubic lattice, with two sublattices ${\mathcal A}$ and
${\mathcal B}$, we get
\begin{eqnarray}
\label{eq:RPAESwmaa}
(\omega-{\omega}^{m}_{\alpha}+{\omega}^{m}_{\alpha'}){G}^{AA}_{\alpha\alpha', \beta\beta'}(\mathbf{q},\omega)
+{P}^{m}_{\alpha\alpha'}\sum_{\mu\nu}T^m_{\alpha\alpha',\nu\mu}(\mathbf{q})G^{BA}_{\mu\nu, \beta\beta'}(\mathbf{q},\omega)=\frac{1}{2\pi}{P}^{m}_{\alpha\alpha'}\delta_{\alpha\beta'}\delta_{\beta\alpha'},
\end{eqnarray}

\begin{eqnarray}
\label{eq:RPAESwmba}
\hspace{-2cm}(\omega-{\omega}^{m}_{\alpha}+{\omega}^{m}_{\alpha'}){G}^{BA}_{\alpha\alpha', \beta\beta'}(\mathbf{q},\omega)
+P^{m}_{\alpha\alpha'}\sum_{\mu\nu}T^m_{\alpha\alpha',\nu\mu}(\mathbf{q})G^{AA}_{\mu\nu, \beta\beta'}(\mathbf{q},\omega) = 0,
 \end{eqnarray}
\end{widetext}
where $T^m_{\alpha \alpha',\beta \beta'}(\textbf{q})=\epsilon_\textbf{q} (T^{ij,m}_{\alpha \alpha',\beta \beta'}+
T^{ji,m}_{\beta \beta',\alpha \alpha'})$, $\epsilon_{q}\equiv-2t\sum_{j=x,y}cos(q_{j})$, ${P}^{m}_{\alpha\alpha'} = \langle {L}^{m}_{\alpha\alpha}\rangle - \langle
{L}^{m}_{\alpha'\alpha'}\rangle$  and
${\omega}^{m}_{\alpha}={V}^{m}_{\alpha\alpha}+\sum_{\beta}{T}^{m}_{\alpha\alpha,\beta\beta}(q=0)\langle
{L}^{m}_{\beta\beta}\rangle$, and $\omega$ and $\mathbf{q}$ are, respectively,
the frequency and wave vector of the excitation. The poles of these Green
functions give the different branches of the excitation spectrum in the RPA.

We can relate the Green functions calculated above to the single-particle Green function $g_{i,j}(t)=-i\theta(t)\langle[\hat{a}_i(t),\hat{a}^{\dagger}_{j}]\rangle$. We first illustrate this for the simple BH model by following the treatment of Ref.~\cite{Matsumoto}. We define the single-particle Green functions
\begin{eqnarray}
\label{eq:bh2}
g_{1}(\textbf{q},\omega)&=&\sum_{\alpha\alpha',~\beta\beta'}y^\dagger_{\alpha,\alpha'}~y_{\beta,\beta'}
G_{\alpha\alpha', \beta \beta'}(\textbf{q},\omega),\\
g_{2}(\textbf{q},\omega)&=&\sum_{\alpha\alpha',~\beta\beta'}y^\dagger_{\alpha,\alpha'}~y^\dagger_{\beta,\beta'}
G_{\alpha\alpha',\beta\beta'}(\textbf{q},\omega),
\end{eqnarray}
 where ${y_{\alpha,\alpha'}~\equiv~\langle i,\alpha |\hat{a}_i| i,\alpha'\rangle }$ and ${y^\dagger_{\alpha,\alpha'}~\equiv~\langle i,\alpha |\hat{a}^\dagger_i| i,\alpha'\rangle }$ are independent of the position
 $i$. It is easy to show that equation of motion for the green functions $g_{1}(\textbf{q},\omega)$ and $g_{2}(\textbf{q},\omega)$ can be written as
\begin{widetext}
\begin{eqnarray}
\label{eq:bh1}
\label{eq:japbh}
\left(
\begin{array}{cccccccc}
\mbox{${1-\epsilon_{q}A_{11}}$} & \mbox{${\epsilon_{q}A_{12}}$} \\
\mbox{${\epsilon_{q}A_{21}}$} & \mbox{${1-\epsilon_{q}A_{22}}$} \\
\end{array}
\right)~\left(
\begin{array}{cccccccc}
\mbox{${g_{1}(q,\omega)}$} \\
\mbox{${g_{2}(q,\omega)}$} \\
\end{array}
\right)~=~\left(
\begin{array}{cccccccc}
\mbox{${A_{11}}$} \\
\mbox{${A_{21}}$} \\
\end{array}
\right),
\end{eqnarray}

where

\begin{eqnarray}
\label{eq:bh4}
\left(
\begin{array}{cccccccc}
\mbox{${A_{11}}$} & \mbox{${A_{12}}$} \\
\mbox{${A_{21}}$} & \mbox{${A_{22}}$} \\
\end{array}
\right)~=~\sum_{\alpha}\frac{\left(
\begin{array}{cccccccc}
\mbox{${{y}_{0,\alpha}{y}^{\dagger}_{\alpha,0}}$} & \mbox{${{y}_{0,\alpha}{y}_{\alpha,0}}$} \\
\mbox{${{y}^{\dagger}_{0,\alpha}{y}^{\dagger}_{\alpha,0}}$} & \mbox{${{y}^{\dagger}_{0,\alpha}{y}_{\alpha,0}}$} \\
\end{array} \right)}{\omega+i\delta+(\omega_{0}-\omega_{\alpha})}~-~\sum_{\alpha}\frac{\left(
\begin{array}{cccccccc}
\mbox{${{y}^{\dagger}_{0,\alpha}{y}_{\alpha,0}}$} & \mbox{${{y}_{0,\alpha}{y}_{\alpha,0}}$} \\
\mbox{${{y}^{\dagger}_{0,\alpha}{y}^{\dagger}_{\alpha,0}}$} & \mbox{${{y}_{0,\alpha}{y}^{\dagger}_{\alpha,0}}$} \\
\end{array} \right)}{\omega+i\delta+(\omega_{\alpha}-\omega_{0})},
\end{eqnarray}
\end{widetext}
$|0\rangle$ represents the ground state and the summation is over all the
excited states $\alpha$.

We give below the analogs of Eqs.(\ref{eq:bh1})-(\ref{eq:bh4}) for
the BH model with two species of bosons in Eqs.(\ref{eq:2bh1}) and (\ref{eq:2bh2}) and for the spin-1 BH model in Eq.(\ref{eq:spin11}).
For $a$-type bosons, we define the single-particle Green functions
\begin{eqnarray}
\label{eq:2bh3}
\nonumber g^{MN}_{1}(\textbf{q},\omega)&=&\sum_{\alpha\alpha',~\beta\beta'}{y^\dagger}^M_{\alpha,\alpha'}~y^N_{\beta,\beta'}
G_{\alpha\alpha', \beta \beta'}(\textbf{q},\omega),\\
&&\\
\nonumber g^{MN}_{2}(\textbf{q},\omega)&=&\sum_{\alpha\alpha',~\beta\beta'}{y^\dagger}^M_{\alpha,\alpha'}~{y^\dagger}^N_{\beta,\beta'}
G_{\alpha\alpha',\beta\beta'}(\textbf{q},\omega),\\
&&
\end{eqnarray}
where the superscripts $M$ and $N$ can be $a$ or $b$, for the two-species BH model, or $\sigma$ and $\sigma'$ for the spin-1 BH model. For the two-species BH model, the equation of motion for the Green functions $g^{MN}_{1}(\textbf{q},\omega)$ and $g^{MN}_{2}(\textbf{q},\omega)$ can be written as

\begin{widetext}
\begin{eqnarray}
\label{eq:2bh1}
\left(
\begin{array}{cccccccc}
\mbox{${1-\epsilon_{p}A^{aa}_{11}}$} & \mbox{${\epsilon_{p}A^{ba}_{11}}$} & \mbox{${\epsilon_{p}A^{aa}_{12}}$} & \mbox{${\epsilon_{p}A^{ba}_{12}}$}\\
\mbox{${\epsilon_{p}A^{ab}_{11}}$} & \mbox{${1-\epsilon_{p}A^{bb}_{11}}$} & \mbox{${\epsilon_{p}A^{ab}_{12}}$} & \mbox{${\epsilon_{p}A^{bb}_{12}}$}\\
\mbox{${\epsilon_{p}A^{aa}_{21}}$} & \mbox{${\epsilon_{p}A^{ba}_{21}}$} & \mbox{${1-\epsilon_{p}A^{aa}_{22}}$} & \mbox{${\epsilon_{p}A^{ba}_{22}}$}\\
\mbox{${\epsilon_{p}A^{ab}_{21}}$} & \mbox{${\epsilon_{p}A^{bb}_{21}}$} & \mbox{${\epsilon_{p}A^{ab}_{22}}$} & \mbox{${1-\epsilon_{p}A^{bb}_{22}}$}
\end{array}
 \right)~\left(
\begin{array}{cccccccc}
\mbox{${g^{aa}_{1}(q,\omega)}$} \\
\mbox{${g^{ba}_{1}(q,\omega)}$} \\
\mbox{${g^{aa}_{2}(q,\omega)}$} \\
\mbox{${g^{ba}_{2}(q,\omega)}$} \\
\end{array}
\right)~=~\left(
\begin{array}{cccccccc}
\mbox{${A^{aa}_{11}}$} \\
\mbox{${A^{ab}_{11}}$} \\
\mbox{${A^{aa}_{21}}$} \\
\mbox{${A^{ab}_{21}}$} \\
\end{array}
\right),
\end{eqnarray}

\begin{eqnarray}
\label{eq:2bh2}
\left(
\begin{array}{cccccccc}
\mbox{${1-\epsilon_{p}A^{bb}_{11}}$} & \mbox{${\epsilon_{p}A^{ab}_{11}}$} & \mbox{${\epsilon_{p}A^{bb}_{12}}$} & \mbox{${\epsilon_{p}A^{ab}_{12}}$}\\
\mbox{${\epsilon_{p}A^{ba}_{11}}$} & \mbox{${1-\epsilon_{p}A^{aa}_{11}}$} & \mbox{${\epsilon_{p}A^{ba}_{12}}$} & \mbox{${\epsilon_{p}A^{aa}_{12}}$}\\
\mbox{${\epsilon_{p}A^{bb}_{21}}$} & \mbox{${\epsilon_{p}A^{ab}_{21}}$} & \mbox{${1-\epsilon_{p}A^{bb}_{22}}$} & \mbox{${\epsilon_{p}A^{ab}_{22}}$}\\
\mbox{${\epsilon_{p}A^{ba}_{21}}$} & \mbox{${\epsilon_{p}A^{aa}_{21}}$} & \mbox{${\epsilon_{p}A^{ba}_{22}}$} & \mbox{${1-\epsilon_{p}A^{aa}_{22}}$}
\end{array}
 \right)~\left(
\begin{array}{cccccccc}
\mbox{${g^{bb}_{1}(q,\omega)}$} \\
\mbox{${g^{ab}_{1}(q,\omega)}$} \\
\mbox{${g^{bb}_{2}(q,\omega)}$} \\
\mbox{${g^{ab}_{2}(q,\omega)}$} \\
\end{array}
\right)~=~\left(
\begin{array}{cccccccc}
\mbox{${A^{bb}_{11}}$} \\
\mbox{${A^{ba}_{11}}$} \\
\mbox{${A^{bb}_{21}}$} \\
\mbox{${A^{ba}_{21}}$} \\
\end{array}
\right).
\end{eqnarray}

For the spin-1 BH model we have:

\begin{equation}
\label{eq:spin11}
\left(
\begin{array}{cccccccc}
\mbox{${1-\epsilon_{p}A^{\sigma\sigma}_{11}}$} & \mbox{${\epsilon_{p}\sum_{\sigma'}A^{\sigma\sigma'}_{12}}$} \\
\mbox{${\epsilon_{p}A^{\sigma'\sigma}_{21}}$} & \mbox{${1-\epsilon_{p}A^{\sigma'\sigma'}_{22}}$} \\
\end{array}
\right)~\left(
\begin{array}{cccccccc}
\mbox{${g^{\sigma\sigma}_{1}(q,\omega)}$} \\
\mbox{${g^{\sigma'\sigma}_{2}(q,\omega)}$} \\
\end{array}
\right)~=~\left(
\begin{array}{cccccccc}
\mbox{${A^{\sigma\sigma}_{11}}$} \\
\mbox{${A^{\sigma'\sigma}_{21}}$} \\
\end{array}
\right).
\end{equation}

Finally we can write

\begin{equation}
\left(
\begin{array}{cccccccc}
\mbox{${A^{MN}_{11}}$} & \mbox{${A^{MN}_{12}}$} \\
\mbox{${A^{MN}_{21}}$} & \mbox{${A^{MN}_{22}}$} \\
\end{array}
\right)~=~\sum_{\alpha}\frac{\left(
\begin{array}{cccccccc}
\mbox{${{y}^{N}_{0,\alpha}{y^\dagger}^{M}_{\alpha,0}}$} & \mbox{${{y}^{N}_{0,\alpha}{y}^{M}_{\alpha,0}}$} \\
\mbox{${{y^\dagger}^{N}_{0,\alpha}{y^\dagger}^{M}_{\alpha,0}}$} & \mbox{${{y^\dagger}^{N}_{0,\alpha}{y}^{M}_{\alpha,0}}$} \\
\end{array} \right)}{\omega+i\delta+(\omega_{0}-\omega_{\alpha})}~-~\sum_{\alpha}\frac{\left(
\begin{array}{cccccccc}
\mbox{${{y^\dagger}^{M}_{0,\alpha}{y}^{N}_{\alpha,0}}$} & \mbox{${{y}^{M}_{0,\alpha}{y}^{N}_{\alpha,0}}$} \\
\mbox{${{y^\dagger}^{M}_{0,\alpha}{y^\dagger}^{N}_{\alpha,0}}$} & \mbox{${{y}^{M}_{0,\alpha}{y^\dagger}^{N}_{\alpha,0}}$} \\
\end{array} \right)}{\omega+i\delta+(\omega_{\alpha}-\omega_{0})}.
\end{equation}

\end{widetext}


\section{Calculation of excitation spectra analytically  in the MI phases for BH models:}

We calculate the excitation spectra for the MI phases of BH models analytically. We begin with the BH model. In the MI phase, the superfluid order parameter is zero, the density of bosons $\rho$ is equal to an integer, and the eigenstates of
the mean-field Hamiltonian are given by the Fock states, $|i,\alpha\rangle\equiv |i,n\rangle =\frac{1}{\sqrt{n!}}(a_i^\dagger)^n|\mbox{vacuum}\rangle$. The presence of $P_{\alpha \alpha'}\equiv \langle L_{\alpha\alpha}\rangle-\langle L_{\alpha'\alpha'}\rangle$ in the RPA equation for the Green function (we have suppressed the site index $i$)

\begin{eqnarray}
\label{eq:RPAESBHw}
(\omega-\omega_{\alpha}+\omega_{\alpha'})&&G_{\alpha\alpha',
\beta\beta'}(\mathbf{q},\omega)+ \\ \nonumber
&&P_{\alpha\alpha'}\sum_{\mu\nu}T_{\alpha\alpha',\nu\mu}(\mathbf{q})G_{\mu\nu,
\beta\beta'}(\mathbf{q},\omega)\\ \nonumber
&&=\frac{1}{2\pi}P_{\alpha\alpha'}\delta_{\alpha\beta'}\delta_{\beta\alpha'}
\end{eqnarray}
yield nonzero Green function $G_{\alpha\alpha',
\beta\beta'}(\mathbf{q},\omega)$ only if $\alpha$ or $\alpha'$ are ground states. Let us obtain the Green function for the ground state with density of bosons equal to $\rho$, $G_{\alpha\alpha',
\alpha'\alpha}(\mathbf{q},\omega)\equiv G_{nn+1,
n+1n}(\mathbf{q},\omega)$ with $|\alpha\rangle=|n=\rho\rangle$ and $|\alpha'\rangle=|n+1\rangle$. The hopping matrix in Eq.~(\ref{eq:RPAESBHw}) is given by
\begin{equation}\label{eq:bhb1}
    T_{\alpha\alpha',\nu\mu}=\left( \langle\alpha| a^\dagger |\alpha'\rangle \langle\nu|a|\mu\rangle+\langle\alpha| a |\alpha'\rangle \langle\nu|a^\dagger|\mu\rangle \right).
\end{equation}
Since $|\alpha\rangle=|n=\rho\rangle$ and $|\alpha'\rangle=|n+1\rangle$, the nonzero term in the above equation have $|\mu\rangle=|n\rangle$, $|\nu\rangle=|n+1\rangle$ or $|\mu\rangle=|n-1\rangle$, $|\nu\rangle=|n\rangle$.
Thus the RPA equation for the Green function $G_{n~n+1,n+1~n}(\mathbf{q},\omega)$ is
\begin{eqnarray}\label{eq:bhb2}
  [\omega-\omega_n+&&\omega_{n+1}+(n+1)\epsilon_q]G_{nn+1,n+1n}(\mathbf{q},\omega)+\nonumber \\ &&\epsilon_q\sqrt{n(n+1)}
  G_{n-1~n,n+1~n}(\mathbf{q},\omega)=\frac{1}{2\pi}.
\end{eqnarray}
Similarly the RPA equation for the Green function $G_{n-1~n,n+1~n}(\mathbf{q},\omega)$ is
\begin{eqnarray}\label{eq:bhb3}
  &&-\epsilon_q\sqrt{n(n+1)}G_{nn+1,n+1n}(\mathbf{q},\omega)\nonumber \\
  &&+
  [\omega-\omega_{n-1}+\omega_{n}-n\epsilon_q]G_{n-1~n,n+1~n}(\mathbf{q},\omega)=0.\nonumber \\
  &&
\end{eqnarray}
For the MI phase, $\omega_n=\mu n-\frac{U}{2}n(n-1)$, so by solving the coupled Eqs. (\ref{eq:bhb2}) and (\ref{eq:bhb3})
we obtain the following excitation spectra for the MI phase with density equal to $\rho$:
\begin{eqnarray}\label{eq:bhb4}
\omega^{\pm}&=&\frac{1}{2}[-(2\mu-U(2\rho-1)+\epsilon_{q})\nonumber \\
&&\pm\sqrt{\epsilon_{q}^{2}-\epsilon_{q}U(4\rho+2)+U^2}];
\end{eqnarray}
here the plus (minus) sign corresponds to the particle (hole) excitation.

The MI phases can also display dispersionless excitation spectra; we
illustrate this by evaluating the Green function
$G_{\alpha\alpha',
\alpha'\alpha}(\mathbf{q},\omega)\equiv G_{n~n+2,
n+2~n}(\mathbf{q},\omega)$ with $|\alpha\rangle=|n=\rho\rangle$ and $|\alpha'\rangle=|n+2\rangle$. It is easily seen from Eq.~(\ref{eq:bhb1}) that,  for these values of $\alpha$ and $\alpha'$, $T_{\alpha\alpha',\nu\mu}(\mathbf{q})=0$ for all values of $\mu$ and $\nu$. Thus the RPA equation for $G_{n~n+2,
n+2~ n}(\mathbf{q},\omega)$ is given by
\begin{equation}\label{eq:bhb5}
    (\omega-\omega_n+\omega_{n+2})G_{n~n+2,n+2~ n}(\mathbf{q},\omega)=\frac{1}{2\pi}.
\end{equation}
This equation gives $\textbf{q}$-independent excitation spectra $\omega=\omega_n-\omega_{n+2}$, i.e., multi-particle (or multi-hole) excitation spectra can be dispersionless.

For the BH model with two-species of bosons, the eigenstates of the mean-field
Hamiltonian for the MI phase, with even total density of bosons $\rho=n_a+n_b$ are given by
\begin{equation}\label{eq:2bhb1}
 |i\alpha\rangle \equiv |i~n_a~n_b\rangle = \frac{1}{\sqrt{n_a!~n_b!}}(a^\dagger)^{n_a}(b^\dagger)^{n_b}|\mbox{vacuum}\rangle.
\end{equation}
 To obtain the particle(hole) excitation for the $a$-type bosons in the MI phase with density of bosons for $a$-type and $b$-type, respectively, equal to $\rho_a$ and $\rho_b$, we calculate the Green function
$G_{\alpha\alpha',\alpha'\alpha}(\mathbf{q},\omega)\equiv G^a_{n_an_a+1}(\mathbf{q},\omega)$ with $|\alpha\rangle=|n_a=\rho_a ~ n_b=\rho_b \rangle$
 and $|\alpha'\rangle=|n_a+1 ~ n_b\rangle$ using the Eq.~(\ref{eq:RPAES2BHw}). In this case, the hopping matrix element $T_{\alpha \alpha',\nu \mu}(\mathbf{q})$ given by Eq.~(\ref{eq:RPAES2BHt}) is non-zero if $\{|\mu\rangle,|\nu\rangle\}$ equal to $\{|n_a~ n_b\rangle,|n_a+1~ n_b\rangle\}$ or $\{|n_a-1~ n_b\rangle,|n_a~ n_b\rangle\}$. Therefore, the RPA equation for $G^a_{n_a~n_a+1}(\mathbf{q},\omega)$ is given by
\begin{eqnarray}\label{eq:2bhb2}
  [\omega-\omega_{n_a~n_b}+&&\omega_{n_a+1~n_b}+(n_a+1)\epsilon_q]G^a_{n_an_a+1}(\mathbf{q},\omega)+\nonumber \\ &&\epsilon_q\sqrt{n_a(n_a+1)}
  G^a_{n_a-1~n_a}(\mathbf{q},\omega)=\frac{1}{2\pi}
\end{eqnarray}
 where the Green function $G^a_{n_a-1~n_a}(\mathbf{q},\omega)\equiv G_{\alpha\alpha',\beta'\beta}(\mathbf{q},\omega)$ with $|\alpha\rangle=|n_a-1~n_b\rangle$, $|\alpha'\rangle=|n_a~n_b\rangle$, $|\beta\rangle=|n_a+1~n_b\rangle$ and $|\beta'\rangle=|n_a~n_b\rangle$ and its RPA equation is given by
 \begin{eqnarray}\label{eq:2bhb3}
  &&-\epsilon_q\sqrt{n_a(n_a+1)}G^a_{n_a~n_a+1}(\mathbf{q},\omega)\nonumber \\
  &&+[\omega-\omega_{n_a-1~n_b}+\omega_{n_a~n_b}-n_a\epsilon_q]G^a_{n_a-1~n_a}(\mathbf{q},\omega)=0.\nonumber \\
  &&~~
\end{eqnarray}
Here $\omega_{n_an_b}=\mu_a n_a-\frac{U_a}{2}n_a(n_a-1)+\mu_b n_b-\frac{U_b}{2}n_b(n_b-1)-U_{ab}n_an_b$.
By solving coupled Eqs. (\ref{eq:2bhb2}) and (\ref{eq:2bhb3}) we obtain the excitation spectra for the MI phase with the densities of $a$-type and $b$-type bosons, respectively, equal to $\rho_a$ and $\rho_b$:
\begin{eqnarray}\label{eq:2bhb4}
\omega^{\pm}&=&\frac{1}{2}[-(2\mu_a-U_a(2\rho_a-1)-2U_{ab}\rho_b+\epsilon_{q})\nonumber \\
&&\pm\sqrt{\epsilon_{q}^{2}-\epsilon_{q}U_a(4\rho_a+2)+U_a^2}]
\end{eqnarray}
where the plus (minus) sign corresponds to the particle (hole) excitation. This expression has been obtained earlier by using Gutzwiller approximation~\cite{ozaki0955}. In the limit $U_{ab}=0$, this expression matches with that for BH model (Eq.~(\ref{eq:bhb4})). Similarly, for the particle (hole) excitations for $b$-type particles we obtain
\begin{eqnarray}\label{eq:2bhb5}
\omega^{\pm}&=&\frac{1}{2}[-(2\mu_b-U_b(2\rho_b-1)-2U_{ab}\rho_a+\epsilon_{q})\nonumber \\
&&\pm\sqrt{\epsilon_{q}^{2}-\epsilon_{q}U_b(4\rho_b+2)+U_b^2}].
\end{eqnarray}

For the MI phase with total density of bosons $\rho=1$, the first few eigenstates of the mean-field Hamiltonian are given by $|\alpha_0\rangle=C_1|1~0\rangle+C_2|0~1\rangle$,  $|\alpha_1\rangle=C_2|1~0\rangle-C_1|0~1\rangle$,  $|\alpha_2\rangle=|0~0\rangle$,  $|\alpha_3\rangle=|1~1\rangle$,  $|\alpha_4\rangle=D_1|2~0\rangle+D_2|0~2\rangle$, and $|\alpha_5\rangle=D_2|2~0\rangle-D_1|0~2\rangle$. Here $C_1$, $C_2$, $D_1$, and $D_2$ depend on the values of $U_a$, $U_b$ and $U_{ab}$.
     The ground state $|\alpha_0\rangle$ couples to mean-field eigenstates (i) $|\alpha_2\rangle$ via the annihilation of $a$-type or $b$-type bosons (ii) $|\alpha_3\rangle$, $|\alpha_4\rangle$ and $|\alpha_5\rangle$  via the creation of $a$-type or $b$-type bosons. Starting with the Green function $G_{\alpha_0 \alpha_2,\alpha_2 \alpha_0}$ the coupled RPA equations can be written in a matrix form given by
\begin{widetext}
\begin{eqnarray}\label{eq:2bhb6}
&&\left(
\begin{array}{cccc}
\omega-\Delta\omega_{\alpha_0,\alpha_2}+T^{\textbf{q}}_{\alpha_0\alpha_2,\alpha_2\alpha_0}& T^{\textbf{q}}_{\alpha_0\alpha_2,\alpha_0\alpha_3} & T^{\textbf{q}}_{\alpha_0\alpha_2,\alpha_0\alpha_4} & T^{\textbf{q}}_{\alpha_0\alpha_2,\alpha_0\alpha_5} \\
-T^{\textbf{q}}_{\alpha_3\alpha_0,\alpha_2\alpha_0}&\omega-\Delta\omega_{\alpha_3,\alpha_0}-T^{\textbf{q}}_{\alpha_3\alpha_0,\alpha_0\alpha_3} &  -T^{\textbf{q}}_{\alpha_3\alpha_0,\alpha_0\alpha_4} & -T^{\textbf{q}}_{\alpha_3\alpha_0,\alpha_0\alpha_5} \\
-T^{\textbf{q}}_{\alpha_4\alpha_0,\alpha_2\alpha_0} &-T^{\textbf{q}}_{\alpha_4\alpha_0,\alpha_0\alpha_3} &\omega-\Delta\omega_{\alpha_4,\alpha_0}-T^{\textbf{q}}_{\alpha_4\alpha_0,\alpha_0\alpha_4}&  -T^{\textbf{q}}_{\alpha_4\alpha_0,\alpha_0\alpha_5}\\
-T^{\textbf{q}}_{\alpha_5\alpha_0,\alpha_2\alpha_0} &-T^{\textbf{q}}_{\alpha_5\alpha_0,\alpha_0\alpha_3} &-T^{\textbf{q}}_{\alpha_5\alpha_0,\alpha_0\alpha_4}&\omega -\Delta\omega_{\alpha_5,\alpha_0}-T^{\textbf{q}}_{\alpha_5\alpha_0,\alpha_0\alpha_5}  \\
\end{array}
\right)\nonumber \\
&&
\left(
\begin{array}{c}
G_{\alpha_0 \alpha_2,\alpha_2 \alpha_0}(\mathbf{q},\omega)\\
G_{\alpha_3 \alpha_0,\alpha_2 \alpha_0}(\mathbf{q},\omega)\\
G_{\alpha_4 \alpha_0,\alpha_2 \alpha_0}(\mathbf{q},\omega)\\
G_{\alpha_5 \alpha_0,\alpha_2 \alpha_0}(\mathbf{q},\omega)
\end{array}
\right)=
\left(
\begin{array}{c}
\frac{1}{2\pi}\\
0\\
0\\
0
\end{array}
\right),
\end{eqnarray}
\end{widetext}
where $\Delta\omega_{\alpha_i,\alpha_j}=\omega_{\alpha_i}-\omega_{\alpha_j} $ and $T^{\textbf{q}}_{\alpha_i\alpha_j,\alpha_j\alpha_1}=T_{\alpha_i\alpha_j,\alpha_j\alpha_1}(\textbf{q})$.
Solutions of this equation give one-hole and three-particle excitation spectra.

The excitation spectra of the MI phase of the spin-1 BH model can also be obtained in a similar manner. The eigenstates of the mean-field Hamiltonian for the MI phase with total density of bosons $\rho=n_1+n_0+n_{-1}$ are given by
\begin{eqnarray}\label{eq:spin1b1}
\nonumber |i\alpha\rangle& \equiv& |i~n_1~n_0~ n_{-1}\rangle
 = \frac{1}{\sqrt{n_1!n_0!n_{-1}!}}\\ &&(a_{i,1}^\dagger)^{n_1}(a_{i,0}^\dagger)^{n_0}(a_{i,-1}^\dagger)^{n_{-1}}|\mbox{vacuum}\rangle.
\end{eqnarray}
First we discuss the MI phase with $\rho=1$ where we find the ground
state is triply degenerate and the eigenstates are linear
combinations of Fock states $|0~0~1\rangle$, $|0~1~0\rangle$ and
$|1~0~0\rangle$. The first excited state is $|0~0~0\rangle$ and is
non-degenerate and its value is independent of $U_2$. The ground state is coupled to this state via the annihilation of a boson, which
gives rise to a hole excitation in the spectrum. The next excited states have degeneracy of six if $U_2 = 0$; this is partially lifted if $U_2$ is finite. These eigenstates are linear combinations of Fock
states $|1~1~0\rangle$, $|1~0~1\rangle$, $|0~1~1\rangle$,
$|2~0~0\rangle$, $|0~2~0\rangle$ and $|0~0~2\rangle$. The ground state couples to these states via the creation of a boson; this gives rise to six-particle excitations in the spectra. These are the
relevant excitations in the $\rho=1$ MI phase. All the higher energy
eigenstates have total boson number more than two and will not
couple to the ground state through the hopping matrix and thus
gives dispersionless multi-particle excitation spectra.

For $\rho=2$, the ground state of the spin-1 mean-field Hamiltonian is
non-degenerate in the case of $U_2>0$. However, it has a degeneracy
of five and six, respectively, for $U_2<0$ and $U_2=0$. These states are linear combinations of the two-boson Fock states $|0~0~2\rangle$,
$|0~2~0\rangle$, $|2~0~0\rangle$, $|0~1~1\rangle$, $|1~0~1\rangle$
and $|1~1~0\rangle$.
 The ground state couples to the one-boson Fock states $|1~0~0\rangle$, $|0~1~0\rangle$
and $|0~0~1\rangle$ via the hopping matrix;
this yields three-hole excitations. Similarly, the ground state couples to three-boson Fock states (a total of 10 states)
via the creation of bosons. All the higher excitations in the mean-field solution do not couple to the ground state through the hopping matrix; and they give rise to dispersionless multi-particle excitations.


\end{document}